\documentclass[aps,pre,twocolumn,superscriptaddress,10pt]{revtex4}
\usepackage[dvips]{graphics}
\usepackage{graphicx}
\usepackage{amsfonts}
\usepackage{amssymb}
\usepackage{amsmath}
\usepackage{subfigure}

\begin{document}

\title{\textbf{Bright-Dark Soliton Complexes in Spinor Bose-Einstein
Condensates}}
\author{H. E. Nistazakis}
\affiliation{Department of Physics, University of Athens, Panepistimiopolis, Zografos,
Athens 15784, Greece }
\author{D.J.\ Frantzeskakis}
\affiliation{Department of Physics, University of Athens, Panepistimiopolis, Zografos,
Athens 15784, Greece }
\author{P.G.\ Kevrekidis}
\affiliation{Department of Mathematics and Statistics, University of Massachusetts,
Amherst MA 01003-4515, USA}
\author{B.A.\ Malomed}
\affiliation{Department of Interdisciplinary Studies, Faculty of Engineering, Tel Aviv
University, Tel Aviv 69978, Israel}
\author{R.\ Carretero-Gonz\'alez}
\affiliation{Nonlinear Dynamical Systems Group, Department of Mathematics and Statistics,
and Computational Science Research Center, San Diego State University, San
Diego CA, 92182-7720, USA}

\begin{abstract}
We present 
bright-dark vector solitons in
quasi-one-dimensional spinor ($F=1$) Bose-Einstein condensates. Using a
multiscale expansion technique, we reduce the corresponding nonintegrable
system of three coupled Gross-Pitaevskii equations (GPEs) to a completely
integrable Yajima-Oikawa system. In this way, we obtain approximate
solutions for small-amplitude vector solitons of dark-dark-bright and
bright-bright-dark types, in terms of the $m_{F}=+1,-1,0$ spinor components,
respectively. By means of numerical simulations of the full GPE system, we
demonstrate that these states indeed feature soliton properties, i.e., they
propagate undistorted and undergo quasi-elastic collisions. It is also shown
that, in the presence of a parabolic trap of strength $\omega $, the bright
component(s) is (are) guided by the dark one(s), and, as a result,  
the small-amplitude vector soliton as a whole performs harmonic 
oscillations of frequency $\omega/ \sqrt{2}$ 
in the shallow soliton limit. We investigate numerically deviations
from this prediction, as the depth of the solitons is increased, as well as
when the strength of the spin-dependent interaction is modified.
\end{abstract}

\date{Submitted to {\em Phys.~Rev.~A, May 2007}}

\maketitle

\section{Introduction}

The development of far-off-resonant optical techniques for trapping of
ultracold atomic gases has opened new directions in the studies of
Bose-Einstein condensates (BECs), allowing one to confine atoms regardless
of their spin (hyperfine) state, see, e.g., Ref.~\cite{ket0}. One of major
achievements in this direction was the experimental creation of \textit{%
spinor} BECs \cite{ket1,cahn}, in which the spin degree of freedom (frozen
in magnetic traps) comes into play. This gave rise to the observation of
various phenomena that are not present in single-component BECs, including
formation of spin domains \cite{spindomain} and spin textures \cite{spintext}.

A spinor condensate formed by atoms with spin $F$ is described by a 
$(2F+1)$-component macroscopic wave function. 
Accordingly, a number of theoretical
works have been dealing with multi-component (\textit{vector}) solitons in
$F=1$ spinor BECs. Bright \cite{wad1,boris,zh} and dark \cite{wad2} 
solitons, as well as 
gap solitons \cite{ofyspin}, have been predicted in this
context (the latter type requires the presence of an optical lattice).
However, \textit{mixed} vector soliton solutions (in particular, ones
composed of bright and dark components) for the respective system of coupled
Gross-Pitaevskii equations (GPEs) have not been reported yet, to the best of
our knowledge. Actually, compound solitons of the mixed type may be of
particular interest, as they would provide for the possibility of \textit{%
all-matter-wave waveguiding}, with the dark soliton component forming an
effective guide for the bright component, similar to the all-optical
waveguiding studied in detail (chiefly, theoretically) in nonlinear optics
\cite{kiag}. Such waveguides could be useful for applications, such as
quantum switches and splitters emulating their optical counterparts \cite%
{bld}.

On the other hand, mixed solitons of the dark-bright type were considered in
a model of a two-component condensate, described by two coupled GPEs \cite%
{BA}. Actually, the model also assumed that the two components represented
different spin states of the same atomic species, with equal scattering
lengths of the intra-component and inter-component atomic collisions (i.e.,
the matrix of the nonlinear coefficients in the coupled GPEs was of the
Manakov's type, that makes the system integrable in the absence of the
external potentials).

In this work we consider a quasi-one-dimensional (quasi-1D) $F=1$ spinor
condensate, described by a system of three coupled GPEs. In the physically
relevant case of $^{87}$Rb and $^{23}$Na atoms with $F=1$, which are known
to form spinor condensates of \textit{ferromagnetic }and \textit{polar}
types, respectively (the definitions are given below), the system includes a
naturally occurring small parameter, $\delta $, namely, the ratio of the
strengths of the spin-dependent and spin-independent interatomic
interactions \cite{kemp,greene}. In the case of $\delta =0$ and without the
external potential, the system of the three coupled GPEs becomes 
the so-called Manakov system \cite{manakov}, 
which is completely integrable \cite{intman}. 
Exploiting the smallness of $\delta $, we will develop a
multiscale-expansion method to asymptotically reduce the
nonintegrable GPE\ system to another \textit{completely
integrable} one, \textit{viz}., the Yajima-Oikawa (YO) system. The latter 
was originally derived to describe the interaction of Langmuir and
sound waves in plasmas \cite{yo} and has been used in studies of
vector solitons in the context of optics \cite{ofyol} and binary
BECs \cite{maximo}. The asymptotic reduction is valid for
homogeneous polar spinor BECs (such as $^{23}$Na), that are not
subject to the modulational instability \cite{ofy2,boris}.
Borrowing exact soliton solutions of the YO system, we predict two
types of vector-soliton complexes in the spinor condensate,
\textit{viz}., dark-dark-bright (DDB) and bright-bright-dark (BBD)
ones for the $m_{F}=+1,-1,0$ spin components. Numerical
simulations of the underlying (full) GPE system show that these
solitary pulses (including ones with moderate, rather than small
amplitudes) emulate genuine solitons (solitary waves in integrable
systems) quite well, propagating undistorted for long times and
undergoing quasi-elastic collisions (quasi-elastic collisions of
solitons in the nonintegrable two-component system were mentioned
earlier in Ref.~\cite{BA}). 

The effect of the harmonic 
trapping potential on the solitons is also studied, analytically
and numerically (the potential also breaks the exact integrability of the coupled GPE equations, even with 
$\delta =0$). It is shown that the small-amplitude vector solitons of the mixed types perform 
harmonic oscillations in the presence of the trap of strength $\omega $, at
frequency $\omega /\sqrt{2}$, which coincides with the well-known frequency
of oscillations of a dark soliton in the single-component BEC \cite{oscfreq}. 
In particular, the bright component(s) of the vector soliton oscillates at
the same frequency, following their dark counterpart(s) (ordinary bright
solitons in single-component BECs oscillate at frequency $\omega $, rather
than $\omega /\sqrt{2}$ \cite{st}, according to the Kohn's theorem \cite{kohn}). 
As a matter of fact, it is a manifestation of the guidance of the
bright component by the dark one. The oscillations of moderate- and large-amplitude solitons are studied 
as well and we find that the oscillation frequency is down-shifted from 
its value $\omega /\sqrt{2}$ as the soliton depth is increased. In the particular 
case of large-amplitude solitons (that perform small-amplitude oscillations around the trap center), 
the frequency down-shift can be very well approximated by the prediction of Ref.~\cite{BA}.

We also investigate the effect of a larger normalized spin-dependent 
interaction strength $\delta$ on the stability of the predicted vector soliton solutions. 
In fact, we consider a value of $\delta$ an order of magnitude larger than the 
physically relevant one of the polar sodium spinor condensate ($\delta = 0.2$  
rather than $0.0314$) and investigate numerically the respective coupled GPEs. 
The result is that, generally, under such a strong perturbation the solitons emit  
stronger radiation and are eventually destroyed. However, even in such a case of a large $\delta$, 
small- and moderate-amplitude DDB solitons are found to persist up to relatively 
large times, of order of more than 300 ms in physical units. Given that for the physically relevant 
small value of $\delta$ pertaining to the sodium spinor BEC, the respective lifetime was 
four times as large, we believe that the vector solitons derived in this work have a good chance to be 
observed experimentally.

The paper is organized as follows: In Sec.~\ref{SEC:model} we present the model, expound our 
analytical approach for the homogeneous system, and derive solutions for the bright-dark 
soliton complexes. Section \ref{SEC:dynamics} is devoted to the presentation 
of the numerical and analytical 
results for the dynamics of the solitons in both
the homogeneous and inhomogeneous (harmonically confined) system. Finally, 
Sec.~\ref{SEC:conclu} concludes the paper.

\section{The model and its analytical consideration\label{SEC:model}}

\subsection{The model\label{SEC:modelA}}

At sufficiently low temperatures (finite temperature effects 
have been considered recently in Refs.~\cite{Guilleumas}),
and in the framework of the mean-field
approach, the spinor BEC\ with $F=1$ is described by the vector order
parameter, $\mathbf{\Psi }(\mathbf{r},t)=[\Psi _{-1}(\mathbf{r},t),\Psi _{0}(%
\mathbf{r},t),\Psi _{+1}(\mathbf{r},t)]^{T}$, with the components
corresponding to the three values of the vertical spin projection, $%
m_{F}=-1,0,+1$. 
Assuming that the condensate is confined in a highly
anisotropic trap with frequencies $\omega _{x}\ll \omega _{\perp }$, we may
assume the wave functions approximately separable, $\Psi _{0,\pm 1}\approx
\psi _{0,\pm 1}(x)\psi _{\perp }(y,z)$, where the transverse wave function $%
\psi _{\perp }(y,z)$ is the ground state of the respective harmonic
oscillator. Then, averaging of the underlying system of the coupled 
three-dimensional (3D) GPEs in transverse plane $(y,z)$ \cite{gpe1d} leads to
the following system of coupled 1D equations for the longitudinal components
of the wave functions (see also Refs.~\cite{wad1,wad2,boris,zh,ofyspin}):
\begin{eqnarray}
i\hbar \partial _{t}\psi _{\pm 1} &\!\!=\!\!&\hat{H}_{\mathrm{si}}\psi _{\pm
1}+c_{2}^{(1D)}(|\psi _{\pm 1}|^{2}+|\psi _{0}|^{2}-|\psi _{\mp 1}|^{2})\psi
_{\pm 1}  \notag \\
&\!\!+\!\!&c_{2}^{(1D)}\psi _{0}^{2}\psi _{\mp 1}^{\ast  },  \label{mgp1} \\[1ex]
i\hbar \partial _{t}\psi _{0} &\!\!=\!\!&\hat{H}_{\mathrm{si}}\psi
_{0}+c_{2}^{(1D)}(|\psi _{-1}|^{2}+|\psi _{+1}|^{2})\psi _{0}  \notag \\
&\!\!+\!\!&2c_{2}^{(1D)}\psi _{-1}\psi _{0}^{\ast  }\psi _{+1},  \label{mgp2}
\end{eqnarray}%
where the asterisk denotes the complex conjugate, and $\hat{H}_{\mathrm{si}%
}\equiv -(\hbar ^{2}/2m)\partial _{x}^{2}+(1/2)m\omega
_{x}^{2}x^{2}+c_{0}^{(1D)}n_{\mathrm{tot}}$ is the spin-independent part of
the Hamiltonian, with $n_{\mathrm{tot}}=|\psi _{-1}|^{2}+|\psi
_{0}|^{2}+|\psi _{+1}|^{2}$ being the total density ($m$ is the atomic
mass). The nonlinearity coefficients have an effectively 1D form, namely $%
c_{0}^{(1D)}=c_{0}/2\pi a_{\perp }^{2}$ and $c_{2}^{(1D)}=c_{2}/2\pi
a_{\perp }^{2}$, where $a_{\perp }=\sqrt{\hbar /m\omega _{\perp }}$ is the
transverse harmonic oscillator length, which defines the size of the
transverse ground state. Finally, coupling constants $c_{0}$ and $c_{2}$
which account, respectively, for spin-independent and spin-dependent
collisions between identical spin-$1$ bosons, are given by (in the
mean-field approximation)
\begin{equation}
c_{0}=\frac{4\pi \hbar ^{2}(a_{0}+2a_{2})}{3m},\,\,\,\,\,\,\,c_{2}=\frac{%
4\pi \hbar ^{2}(a_{2}-a_{0})}{3m},  \label{c0c2}
\end{equation}%
where $a_{0}$ and $a_{2}$ are the $s$-wave scattering lengths in the
symmetric channels with total spin of the colliding atoms $F=0$ and $F=2$,
respectively. Note that the $F=1$ spinor condensate may be either \textit{%
ferromagnetic} (such as the $^{87}$Rb), characterized by $c_{2}<0$, or
\textit{polar} (such as the $^{23}$Na), with $c_{2}>0$ \cite{ho,ohmi}.

Measuring time, length and density in units of $\hbar /c_{0}^{(1D)}n_{0}$, $%
\hbar /\sqrt{mc_{0}^{(1D)}n_{0}}$ and $n_{0}$, respectively (where $n_{0}$
is the peak density), we cast Eqs.~(\ref{mgp1})--(\ref{mgp2}) in the
dimensionless form,
\begin{eqnarray}
i\partial _{t}\psi _{\pm 1} &=&H_{\mathrm{si}}\psi _{\pm 1}+\delta (|\psi
_{\pm 1}|^{2}+|\psi _{0}|^{2}-|\psi _{\mp 1}|^{2})\psi _{\pm 1}  \notag \\
&+&\delta \psi _{0}^{2}\psi _{\mp 1}^{\ast  },  \label{dvgp1} \\[1ex]
i\partial _{t}\psi _{0} &=&H_{\mathrm{si}}\psi _{0}+\delta (|\psi
_{-1}|^{2}+|\psi _{+1}|^{2})\psi _{0}  \notag \\
&+&2\delta \psi _{-1}\psi _{0}^{\ast  }\psi _{+1},  \label{dvgp2}
\end{eqnarray}%
where $H_{\mathrm{si}}\equiv -(1/2)\partial _{x}^{2}+(1/2)\Omega _{\mathrm{tr%
}}^{2}x^{2}+n_{\mathrm{tot}}$, the normalized harmonic trap strength is
given by
\begin{equation}
\Omega _{\mathrm{tr}}=\frac{\omega _{x}}{\omega _{\perp }}\frac{3}{%
2(a_{0}+2a_{2})n_{0}},  \label{Omegatr}  \\[-1.0ex]
\end{equation}%
and we define
\vspace{-0.2cm}
\begin{equation}
\delta \equiv \frac{c_{2}^{(1D)}}{c_{0}^{(1D)}}=\frac{a_{2}-a_{0}}{%
a_{0}+2a_{2}}.  \label{delta}
\end{equation}%
According to what was said above, $\delta <0$ and $\delta >0$ correspond,
respectively, to ferromagnetic and polar spinor BECs. In the relevant cases
of $^{87}$Rb and $^{23}$Na atoms with $F=1$, $\delta =-4.66\times 10^{-3}$
\cite{kemp} and $\delta =+3.14\times 10^{-2}$ \cite{greene}, respectively,
i.e., in either case, $\delta $ plays the role of a small parameter of Eqs.~(%
\ref{dvgp1})--(\ref{dvgp2}).

Generally, Eqs.~(\ref{dvgp1})--(\ref{dvgp2}) give rise to spin-mixing states
\cite{sm}. However, there also exist non-spin-mixing, or spin-polarized
states, which are stable stationary solutions of Eqs.~(\ref{dvgp1}) and (\ref%
{dvgp2}) \cite{ho,yi}. Here, we first consider the spatially homogeneous
system ($\Omega_{\mathrm{tr}} =0$), and focus on such solutions having at least one
component equal to zero, the remaining ones being continuous waves (CWs).
The corresponding exact stationary solutions are
\begin{equation}
\psi _{-1} =\psi _{+1}=\sqrt{\frac{\mu }{2}}\exp (-i\mu t), \quad\psi_{0}=0,  \\[-2.0ex]
\label{cw1}
\end{equation}
and
\vspace{-0.4cm}
\begin{equation}
\psi _{-1} =\psi _{+1}=0,\quad\psi _{0}=\sqrt{\mu }\exp (-i\mu t).
\label{cw2}
\end{equation}
As we demonstrate below, weakly nonlinear perturbations around these
solutions take the form of three-component dark-bright soliton complexes. In
particular, perturbations around solutions (\ref{cw1}) and (\ref{cw2}) will
lead to solitons of the DDB and BBD types, respectively, for components $%
\psi _{\pm 1}$ and $\psi _{0}$. Since the analytical approach and the
derivation of the soliton solutions for the two cases are quite similar, we
focus below on the DDB solitons, and discuss the BBD ones in a brief form.

It is relevant to note that, for solutions with $\psi _{+1}=\psi _{-1}\equiv
\psi _{1}$, two equations (\ref{dvgp1}) coalesce into a single one for wave
function $\psi _{1}$. Then, the transformation $\psi _{1}\equiv \left( \phi
_{D}+\phi _{B}\right) /\left( 2\sqrt{1+\delta ^{2}}\right) $, $\psi
_{0}\equiv \left( \phi _{D}-\phi _{B}\right) /\sqrt{1+\delta ^{2}}$ casts
the system of two equations (\ref{dvgp1}) and (\ref{dvgp2}) into the form of
two coupled GPEs, with nonlinear cross-coupling coefficients $%
g_{D}=g_{B}=\left( 1-\delta \right) /\left( 1+\delta \right) $, which was
introduced in Ref.~\cite{BA} (also with the objective to study bound
complexes of dark-bright solitons, carried by fields $\phi _{D}$ and $\phi
_{B}$, respectively). In fact, the analysis in Ref.~\cite{BA} was limited to
the case of $g_{D}=g_{B}=1$, while in this work we focus on effects
generated by small $\delta \neq 0$. It should also be stressed that,
although stationary equations presented below may indeed be found from the
system of two, rather than three, coupled GPEs, the stability tests for the
solutions are performed against general perturbations (see below), which include those
with $\psi _{+1}\neq \psi _{-1}$ (i.e., the full system of the three
equations was employed in the stability simulations).

\subsection{Linear analysis\label{SEC:modelB}}

Aiming to find solutions of Eqs.~(\ref{dvgp1})--(\ref{dvgp2}) close to the
CW solution given by Eq.~(\ref{cw1}), we start the analysis by adopting the
following ansatz,
\begin{eqnarray}
\psi _{-1}=\psi _{+1} &=&\sqrt{n(x,t)}\exp [-i\mu t+i\phi (x,t)],  \notag \\%
[1ex]
\psi _{0} &=&\Phi _{0}(x,t)\exp (-i\mu t),  \label{ansatz}
\end{eqnarray}%
where $n(x,t)$ and $\phi (x,t)$ are real density and phase of fields $\psi
_{\pm 1}$, while $\Phi _{0}$ is, generally, a complex function. Substituting
Eq.~(\ref{ansatz}) into Eqs.~(\ref{dvgp1})--(\ref{dvgp2}), we derive the
following a system,
\begin{eqnarray}
&&\frac{i}{2}[\partial _{t}n+\partial _{x}(n\partial _{x}\phi )]-n\left[
\partial _{t}\phi +2n-\mu +(1+\delta )|\Phi _{0}|^{2}\right]   \notag \\
&&-n\left[ \frac{1}{2}(\partial _{x}\phi )^{2}-\frac{1}{2\sqrt{n}}\partial
_{x}^{2}\sqrt{n}+\delta \Phi _{0}^{2}e^{-2i\phi }\right] =0,  \label{h1}
\end{eqnarray}%
\begin{eqnarray}
i\partial _{t}\Phi _{0} &+&\frac{1}{2}\partial _{x}^{2}\Phi _{0}-(2n-\mu
+|\Phi _{0}|^{2})\Phi _{0}  \notag \\
&-&2\delta n\left( \Phi _{0}+\Phi _{0}^{\ast  }e^{-2i\phi }\right) =0.
\label{h2}
\end{eqnarray}%
The CW state (\ref{cw1}) corresponds to an obvious solution of Eqs.~(\ref{h1}%
) and (\ref{h2}) with $n=\mu /2$, $\phi =0$, $\Phi _{0}=0$. Next, we
linearize the equations around this state, looking for a solution as $n=(\mu
/2)+\epsilon \tilde{n}$, $\phi =\epsilon \tilde{\phi}$ and $\Phi
_{0}=\epsilon \tilde{\Phi}_{0}$, where $\epsilon $ is a formal small
parameter. At order $O(\epsilon )$, the linearization leads to the following
system:
\begin{eqnarray}
&&i\left( \partial _{t}\tilde{n}+\frac{\mu }{2}\partial _{x}^{2}\tilde{\phi}%
\right) -\mu \left( \partial _{t}\tilde{\phi}+2\tilde{n}-\frac{\mu }{4}%
\partial _{x}^{2}\tilde{n}\right) =0,\qquad   \label{ls1} \\
&&i\partial _{t}\tilde{\Phi}_{0}+\frac{1}{2}\partial _{x}^{2}\tilde{\Phi}%
_{0}-\delta \cdot \mu \left( \tilde{\Phi}_{0}+\tilde{\Phi}_{0}^{\ast 
}\right) =0.  \label{ls2}
\end{eqnarray}%
Combining real and imaginary parts of Eq.~(\ref{ls1}), we arrive at a
dispersive wave equation,
\begin{equation}
\partial _{t}^{2}\tilde{n}-\mu \partial _{x}^{2}\tilde{n}+\left( \mu
^{2}/8\right) \partial _{x}^{4}\tilde{n}=0,  \label{le}
\end{equation}%
which gives rise to a \emph{stable} dispersion relation between wavenumber $k
$ and frequency $\omega $ (the absence of complex roots for $\omega $ at
real $k$ implies the\textit{\ modulational stability }of the underlying CW
state):
\begin{equation}
\omega ^{2}=\mu k^{2}\left( 1+\mu k^{2}/8\right) .  \label{dr1}
\end{equation}%
It follows from Eq.~(\ref{dr1}) that, in the long-wave limit ($k\rightarrow 0
$), small-amplitude waves can propagate on top of CW solution (\ref{cw1})
with the \textit{speed of sound},
\begin{equation}
c=\sqrt{\mu }.  \label{sound}
\end{equation}%
A similar analysis for Eq.~(\ref{ls2}), which is decoupled from Eq.~(\ref%
{ls1}), leads to dispersion relation
\begin{equation}
\omega ^{2}=k^{2}\left( \delta \mu +k^{2}/4\right) .  \label{dr2}
\end{equation}%
It is clear from here that, for $\delta >0$ (which corresponds to the
\textit{polar state}), Eq.~(\ref{dr2}) has no complex roots for $\omega $,
hence the trivial solution to Eq.~(\ref{h2}), $\Phi _{0}=0$, is
modulationally stable. However, $\delta <0$ (corresponding to the \textit{%
ferromagnetic state}) gives rise to modulational instability of the $\Phi
_{0}=0$ solution against the perturbations whose wavenumbers belong to the
instability band, $k\leq 2\sqrt{|\delta |\mu }$. Note that these results
comply with those reported in Ref.~\cite{ofy2}. Below, we focus on the
modulationally stable case, which pertains to the polar state with $\delta >0
$.

\subsection{Asymptotic soliton solutions}

To consider solutions for weakly nonlinear deviations from the CW state, we
recall that $\delta $ is a small parameter of the system (\ref{dvgp1})--(\ref%
{dvgp2}), which suggests to define the stretched variables,
\begin{equation}
X\equiv \delta ^{1/2}(x-\sqrt{\mu }t),\,\,\,\,\,T\equiv \delta t.  \label{sv}
\end{equation}%
Then, we seek for solutions of Eqs.~(\ref{h1})--(\ref{h2}) as
\begin{equation}
n=\left( \mu /2\right) +\delta \cdot \rho ,\,\,\,\,\,\phi =\delta
^{1/2}\alpha ,\,\,\,\,\,\Phi _{0}=\delta ^{3/4}q,  \label{ae}
\end{equation}%
\begin{equation}
q\equiv q_{1}\cos (Kx-\Omega t)+iq_{2}\sin (Kx-\Omega t),  \label{q}
\end{equation}%
where $\rho =\rho (X,T)$, $\alpha =\alpha (X,T)$, $q_{1,2}=q_{1,2}(X,T)$,
while $K$ and $\Omega $ are unknown wavenumber and frequency. Substituting
Eqs.~(\ref{ae}) in Eq.~(\ref{h1}), at the leading order in $\delta $, which
is $O(\delta )$, we derive a relation between density $\rho $ and phase $%
\alpha $,
\begin{equation}
\sqrt{\mu }\partial _{X}\alpha =2\rho .  \label{ar}
\end{equation}%
At the next order $O(\delta ^{3/2})$, the resulting equation is complex:
\begin{equation}
-\left( i\mu /4\right) (2\partial _{X}\rho -\sqrt{\mu }\partial
_{X}^{2}\alpha )+\partial _{T}\alpha +|q|^{2}=0.  \label{i1}
\end{equation}%
The imaginary part of the expression on the left-side of Eq.~(\ref{i1})
vanishes due to the validity of Eq.~(\ref{ar}), while the real part leads to
equation $\partial _{T}\alpha +|q|^{2}=0$. The compatibility condition of
the latter with Eq.~(\ref{ar}) leads to
\begin{equation}
\partial _{T}\rho =-(\sqrt{\mu }/2)\partial _{X}\left( |q|^{2}\right) .
\label{yo1}
\end{equation}

We now proceed to Eq.~(\ref{h2}), which, to leading order in $\delta $,
i.e., at $O(\delta ^{3/4})$, yields the following system:
\begin{eqnarray}
\Omega q_{1}-\left( K^{2}/2\right) q_{2}&=&0,
\nonumber
\\[1.0ex]
-\left[ \left( K^{2}/2\right) +2\mu \right] q_{1}+\Omega q_{2}&=&0.  
\label{ko}
\end{eqnarray}
Nontrivial solutions to Eqs.~(\ref{ko}) are possible when the following
dispersion relation for $\Omega $ and $K$ holds:
\begin{equation}
\Omega ^{2}=K^{2}\left( \mu +K^{2}/4\right) .  \label{dr3}
\end{equation}%
%
%
Next, to order $O(\delta ^{5/4})$, Eq.~(\ref{h2}) leads to system
\begin{eqnarray}
-\sqrt{\mu }\partial _{X}q_{1}+K\partial _{X}q_{2}&=&0,
\nonumber
\\[1.0ex]
-K\partial_{X}q_{1}+\sqrt{\mu }\partial _{X}q_{1}&=&0,
\nonumber
\end{eqnarray}
which has nontrivial solutions if $K^{2}=\mu $.\ In combination with Eq.~(%
\ref{dr3}), the latter relation selects the frequency, $\Omega =5\mu ^{2}/4$%
. Finally, at order $O(\delta ^{7/4})$, Eq.~(\ref{h2}) leads to equation
\begin{equation}
i\partial _{T}q+\frac{1}{2}\partial _{X}^{2}q-2\rho q=0.  \label{yo2}
\end{equation}

Equations (\ref{yo1}) and (\ref{yo2}), which are the basic result of our
analysis, constitute the \textit{Yajima-Oikawa} (YO) system. It describes
the interaction of low-frequency and high-frequency waves, and was
originally derived in the context of plasma physics, where it applies to
Langmuir (high-frequency) waves, forming a packet (soliton) moving at
velocities close to the speed of sound, and thus strongly coupled to the
ion-acoustic (low-frequency) waves \cite{yo}. As shown in Ref.~\cite{yo},
the YO system is integrable by means of the inverse-scattering-transform,
and gives rise to soliton solutions. The solitons have the $-\mathrm{sech}%
^{2}$ shape for field $\rho $, and $\mathrm{sech}$ shape for $q$, which
correspond to a density dip for components $\psi _{\pm 1}$ and a bright
soliton for $\psi _{0}$, as per Eqs.~(\ref{ae}). According to Eq.~(\ref{ar}%
), the phase profile of the $\psi _{\pm 1}$ components, in the form of $%
\tanh $, is associated to the density dip, hence the patterns in these
components, generated by the exact solution of the YO system, are genuine
dark solitons. The full form of the approximate (asymptotic) DDB soliton
solution to Eqs.~(\ref{dvgp1})--(\ref{dvgp2}), into which the YO soliton is
mapped by Eqs.~(\ref{ansatz}), (\ref{sv}), and (\ref{ae}), is
\begin{eqnarray}
&&\psi _{\pm 1}(x,t)=\sqrt{\left( \mu /2\right) -2\delta \eta ^{2}\mathrm{%
sech}^{2}(2\sqrt{\delta }\eta Z)}\qquad   \notag \\[1ex]
&&\times \exp \left[ -i\mu t-2i\eta \sqrt{\delta /\mu }\tanh (2\sqrt{\delta }%
\eta Z)\right] ,  \label{sol1} \\[2ex]
&&\psi _{0}(x,t)=2^{3/2}\delta ^{3/4}\eta \mu ^{-1/4}\sqrt{\xi }\mathrm{sech}%
(2\sqrt{\delta }\eta Z)  \notag \\[1ex]
&&\times \exp \left[ -i\mu t+i\sqrt{\mu }x-2i\sqrt{\delta }\xi Z+2i\delta
(\eta ^{2}-\xi ^{2})t\right] ,\qquad   \label{sol2}
\end{eqnarray}%
where $Z\equiv x-(\sqrt{\mu }-2\sqrt{\delta }\xi )t$, while $\eta $ and $\xi
$ are arbitrary parameters of order $O(1)$.

A similar analysis can be performed to derive asymptotic soliton solutions
of the BBD type. In that case, starting from CW solution (\ref{cw2}), we
seek for solutions of Eqs.~(\ref{dvgp1})--(\ref{dvgp2}) in the form of
\begin{eqnarray}
\psi _{-1} &=&\psi _{+1}=\Phi _{0}(x,t)\exp (-i\mu t),  \notag \\[1ex]
\psi _{0} &=&\sqrt{n(x,t)}\exp [-i\mu t+i\phi (x,t)].  \label{ansatz2}
\end{eqnarray}%
Next, following the same analytical approach which has led above to the DDB
soliton, we again end up with the YO system, in a form similar to Eqs.~(\ref%
{yo1}) and (\ref{yo2}), namely,
\begin{eqnarray}
\partial _{T}\rho =-2\sqrt{\mu }\partial _{X}|q|^{2},
\nonumber
\\[1.0ex]
i\partial _{T}q+\frac{1}{2}\partial _{X}^{2}q-\rho q=0.  \label{yo4}
\end{eqnarray}
Eventually, the approximate BBD soliton solution to Eqs.~(\ref{dvgp1})--(\ref%
{dvgp2}), generated by the YO soliton, is
\begin{eqnarray}
&&\psi _{\pm 1}(x,t)=2\delta ^{3/4}\eta \mu ^{-1/2}\sqrt{\xi }\mathrm{sech}(2%
\sqrt{\delta }\eta Z)  \notag \\[1ex]
&&\times \exp \left[ -i\mu t+i\sqrt{\mu }x-2i\sqrt{\delta }\xi Z+2i\delta
(\eta ^{2}-\xi ^{2})t\right] ,\qquad   \label{sol3} \\[2ex]
&&\psi _{0}(x,t)=\sqrt{(\mu /2)-4\delta \eta ^{2}\mathrm{sech}^{2}(2\sqrt{%
\delta }\eta Z)}  \notag \\[1ex]
&&\times \exp \left[ -i\mu t-2i\eta \sqrt{\delta /\mu }\tanh (2\sqrt{\delta }%
\eta Z)\right] .  \label{sol4}
\end{eqnarray}%
As the latter solution is quite similar to the DDB one, given by Eqs.~(\ref%
{sol1})--(\ref{sol2}), below we only deal with the dynamics of the DDB
solitons. It is worthwhile to note in passing that both types of solutions
are \textit{genuinely traveling} ones, i.e., they do not exist with zero
speed.

\section{Dynamics of DDB spinor solitons\label{SEC:dynamics}}

\subsection{Numerical Results}

In order to test the prediction of the existence of the DDB solitons in the
underlying spinor BEC model, we turn to numerical integration of the
original GPEs (\ref{dvgp1})--(\ref{dvgp2}). In particular, we fix $\delta =0.0314$ (the
value corresponding to $^{23}$Na) and use the following initial conditions
for the densities,
\begin{eqnarray}
|\psi _{\pm 1}(x,t=0)|^{2} &=&\frac{1}{2}\left[ \mu -\nu ~\mathrm{sech}^{2}(%
\sqrt{\nu }x)\right] ,  \label{ic1} \\
|\psi _{0}(x,t=0)|^{2} &=&\frac{\nu ^{3/2}\xi }{\eta \sqrt{\mu }}\mathrm{sech%
}^{2}(\sqrt{\nu }x),  \label{ic2}
\end{eqnarray}%
while initial phase profiles are similar to those in Eqs.~(\ref{sol1})--(\ref%
{sol2}), the parameter that determines the initial width of the soliton
being
\begin{equation}
\nu \equiv 4\eta ^{2}\delta .  \label{nu}
\end{equation}%
Other parameters are chosen as $\mu =2$, $\xi =0.5$, and
$\Omega _{\mathrm{tr}}=0$ (for the homogeneous condensate), or $\Omega _{%
\mathrm{tr}}=0.05$ for the trapped condensate. In physical terms, this
choice corresponds to the spinor condensate of sodium atoms with the peak 1D
density $n_{0}\simeq 10^{8}$ m$^{-1}$, which contains $\simeq 20000$ atoms
confined in the trap with frequencies $\omega _{\perp }=34\omega _{x}=2\pi
\times 230~$Hz; in this case, the time and space units are, respectively, $%
1.2$ ms and $1.8~\mu $m.

Choosing value $\eta =1$ of the arbitrary parameter introduced above, and
substituting $\delta =0.0314$, $\nu =0.13$, we have checked that these
values indeed provide for very good agreement of the analytical predictions
with numerical results. However, we will display numerical results obtained
for an essentially larger value of $\nu $, \textit{viz}., $\nu =1.2$ [which,
as seen from Eq.~(\ref{nu}) corresponds to $\eta =3.091$ and hence, from Eq.~(%
\ref{ic1}), to a soliton complex with deeper and narrower dark components,
and similarly taller and narrower bright components]. In this way, we intend
to showcase the really wide range of validity of the analytical approach,
and the robustness of the obtained solitary wave solutions.

\begin{figure}[tbp]
~~ 
\includegraphics[width=4.10cm,height=2.9cm]{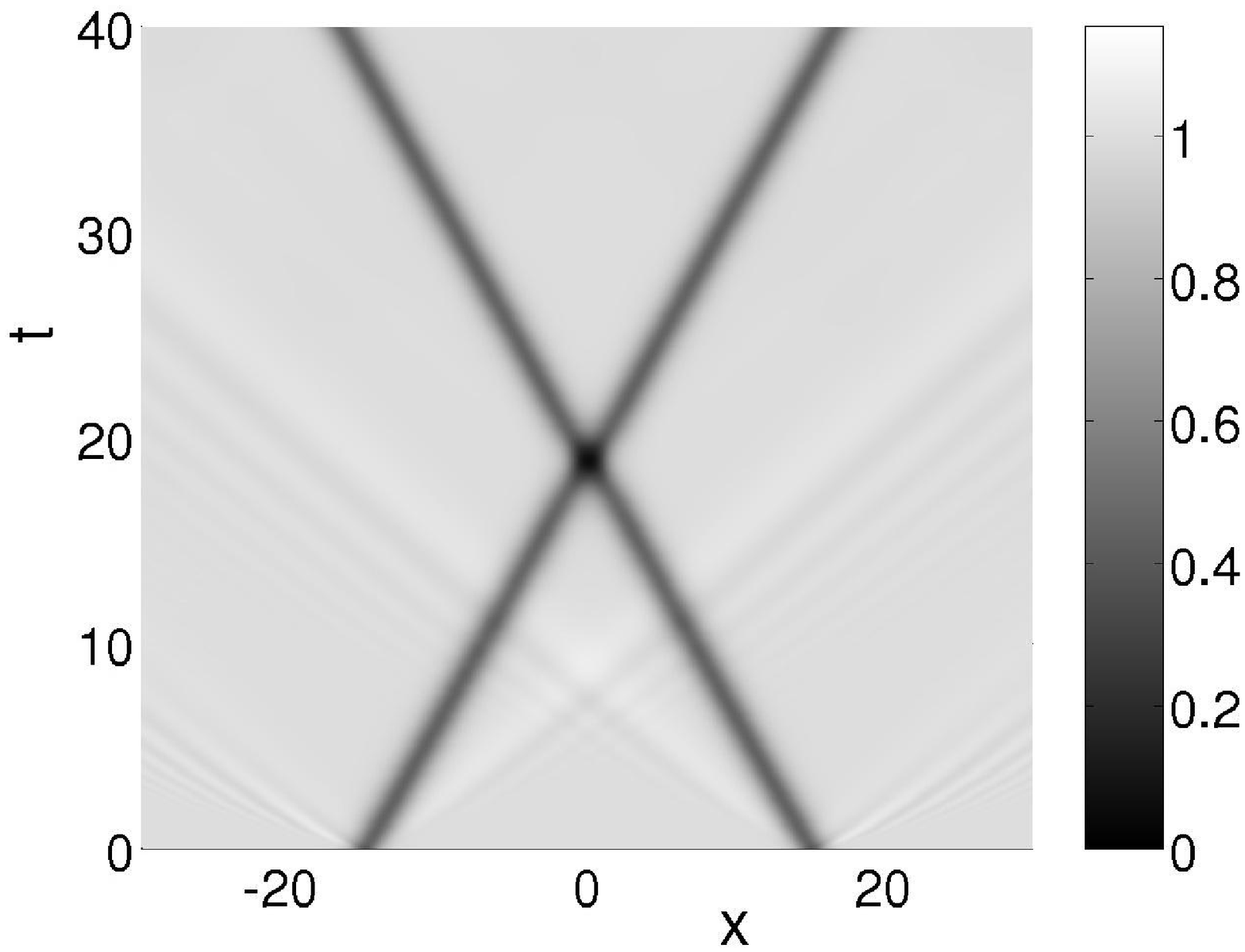} %
\includegraphics[width=4.10cm,height=2.9cm]{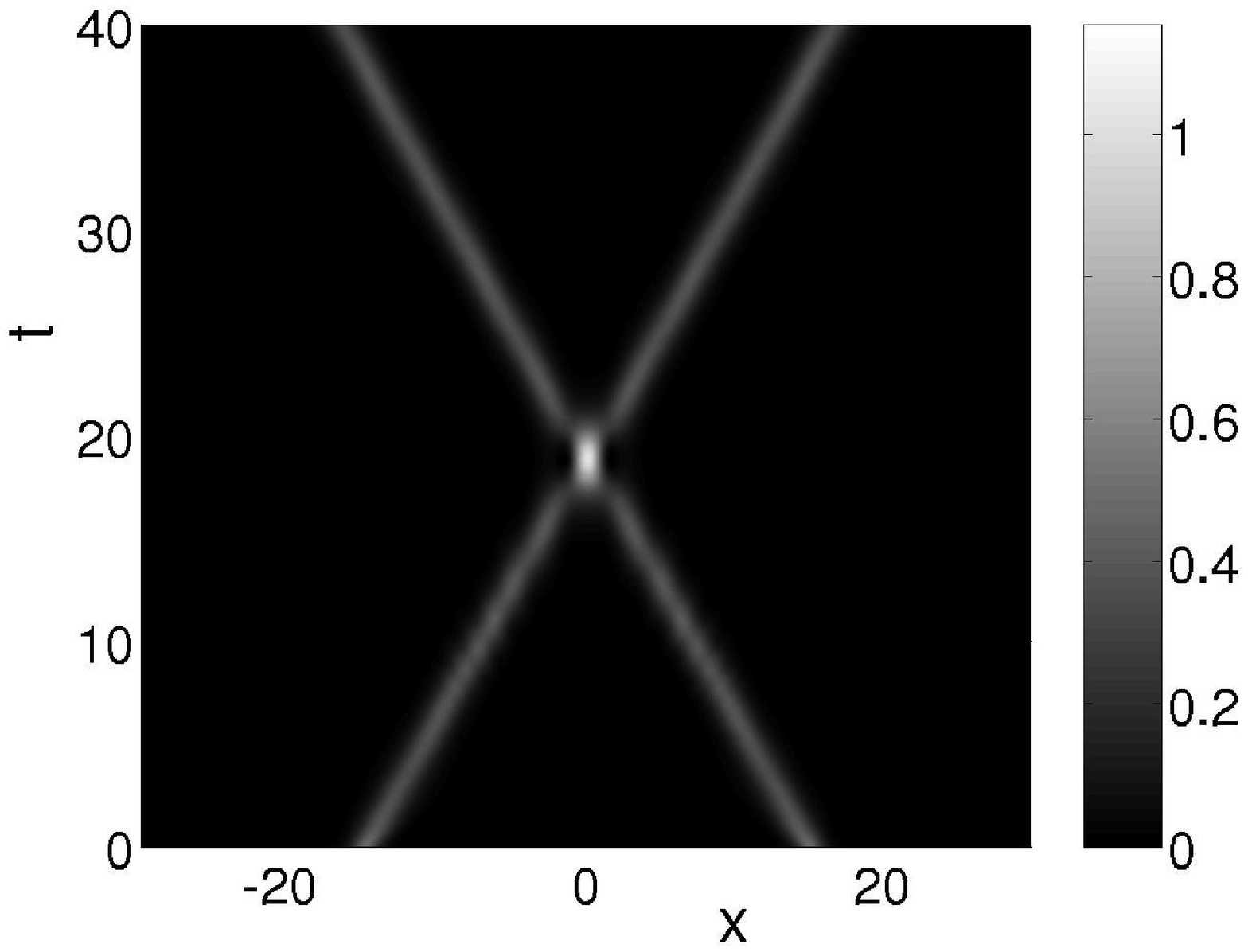}\\[1.0ex]
\includegraphics[width=3.95cm,height=2.9cm]{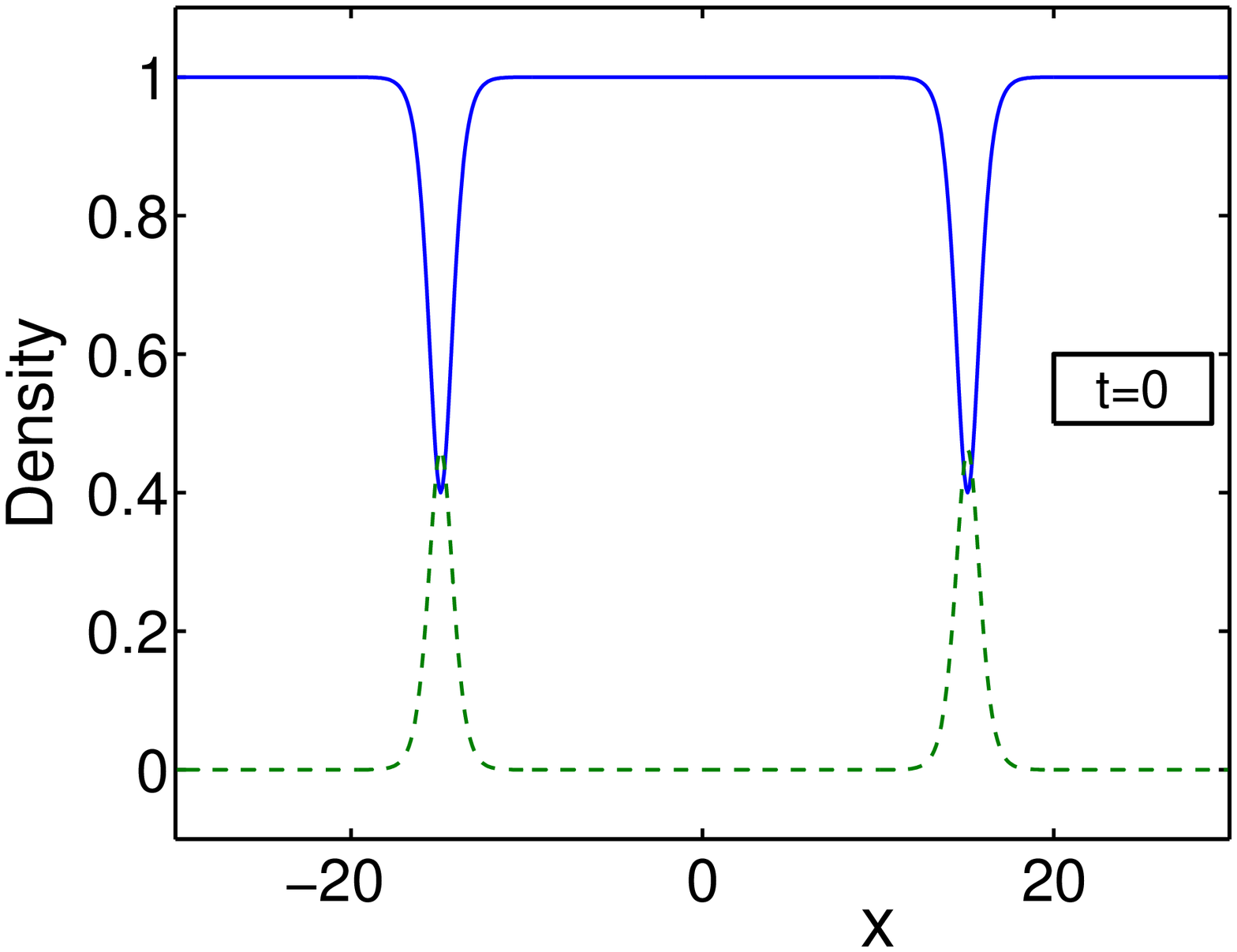} ~ %
\includegraphics[width=3.95cm,height=2.9cm]{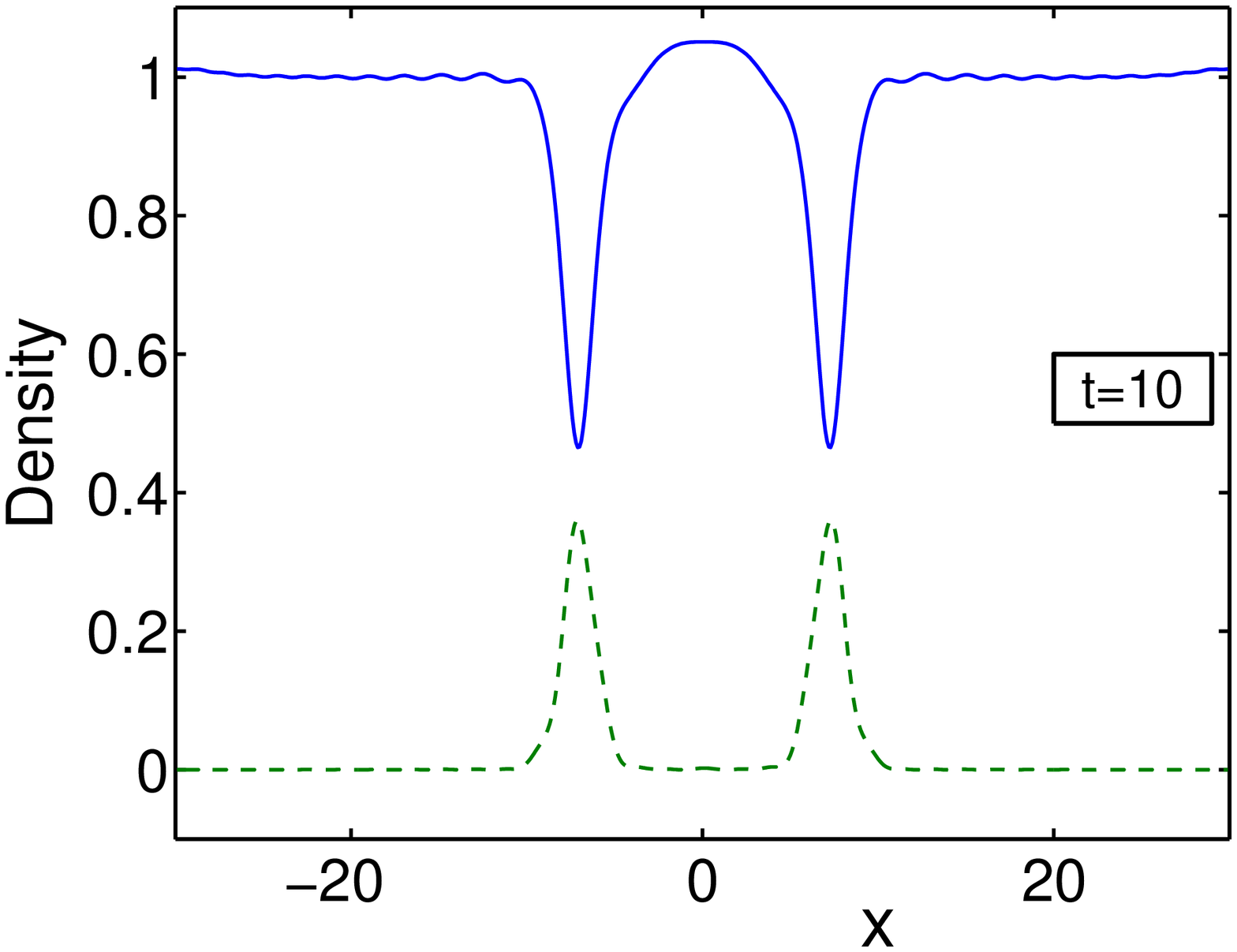}\\[1.0ex]
\includegraphics[width=3.95cm,height=2.9cm]{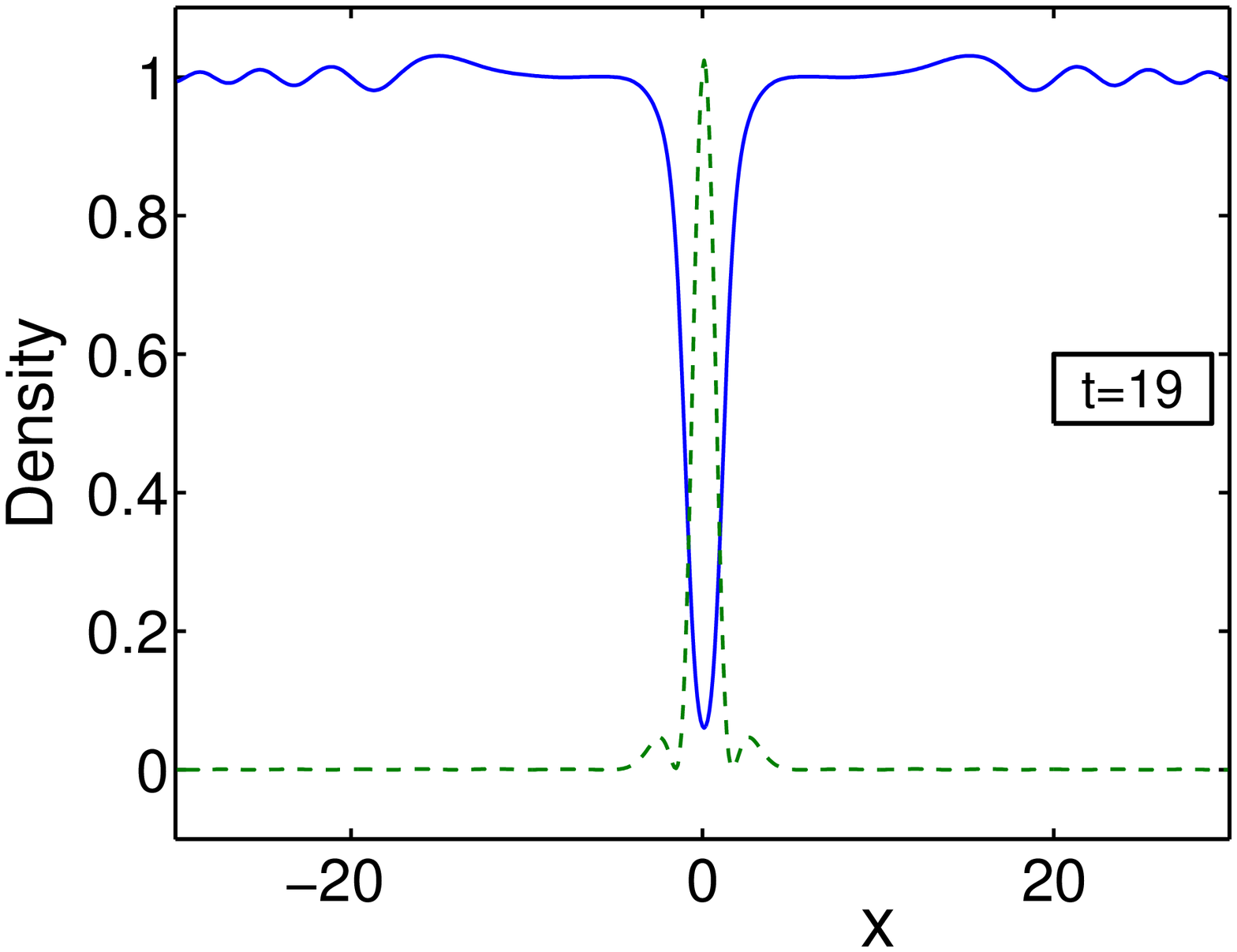} ~ %
\includegraphics[width=3.95cm,height=2.9cm]{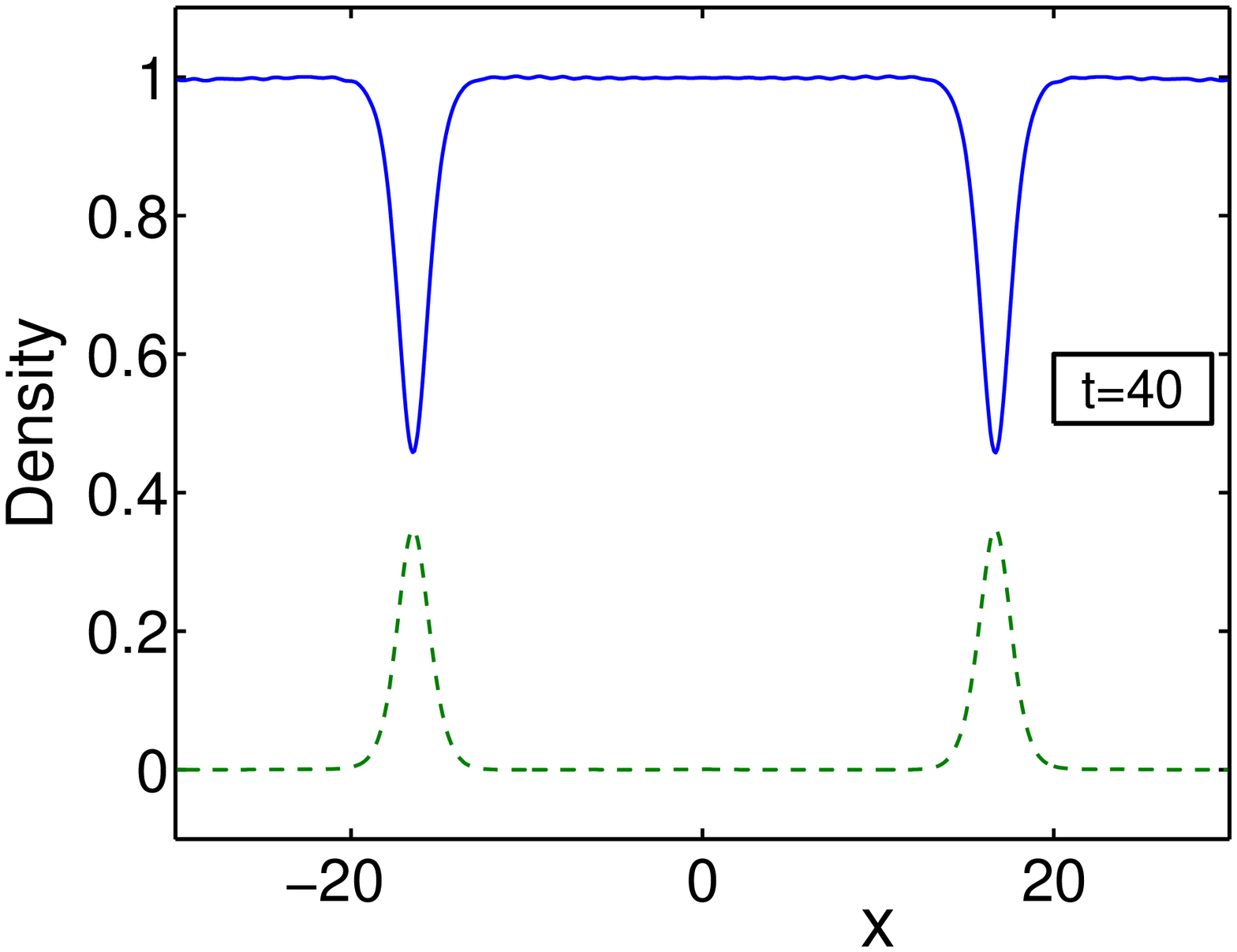}
\vspace{-0.4cm}
\caption{ The two top panels show contour plots of the densities of the $%
\protect\psi _{\pm 1}$ (left panel) and $\protect\psi _{0}$ (right panel)
components of the spinor condensate (with $\protect\delta =0.0314$) in the
homogeneous system ($\Omega _{\mathrm{tr}}=0$). The $\protect\psi _{\pm 1}$
components contain a pair of dark pulses, initially placed at $x_{0}=\pm 15$,
that, together with the bright components in the $\protect\psi _{0}$ field
coupled to them, undergo a quasi-elastic collision at $t\approx 19$ and
propagate unscathed afterward. The parameters are $\protect\mu =2$, $\protect%
\xi =1.54$, $\protect\eta =3.091$, and $\protect\nu =1.2$. The four bottom
panels show snapshots of the densities observed at $t=0,10$ (before the
collision), $t=19$ (when the collision occurs) and $40$ (after the
collision). }
\label{fig1}
\end{figure}

More specifically, we first check if the spinor DDB soliton complexes indeed
behave as genuine solitons, in the small-amplitude limit. To this end, we
take initial conditions in the form of the superposition of two different
pulses,
\begin{eqnarray}
&&\psi _{\pm 1}(x,0) =\sqrt{\frac{\mu }{2}-\frac{\nu }{2}\left[ \mathrm{sech}%
^{2}(x_{+})+\mathrm{sech}^{2}(x_{-})\right] }  
\notag \\
&&\times \exp \left( -i\sqrt{\frac{\nu }{\mu }}\tanh (x_{+})+i\sqrt{\frac{%
\nu }{\mu }}\tanh (x_{-})\right),   
\label{icc1}
\end{eqnarray}%
\vspace{-0.4cm}
\begin{eqnarray}
\psi _{0}(x,0) &=&\nu ^{3/4}\sqrt{\frac{\xi }{\eta \sqrt{\mu }}}\left[
\mathrm{sech}(x_{+})e^{+i\sqrt{\frac{\mu }{\nu }}x_{+}-i(\xi /\eta
)x_{+}}\right.   \notag \\
&&\left. +\mathrm{sech}(x_{-})e^{-i\sqrt{\frac{\mu }{\nu }}x_{-}+i(\xi /\eta
)x_{-}}\right] ,  \label{icc2}
\end{eqnarray}%
where $x_{\pm }=\sqrt{\nu }\left( x\pm x_{0}\right) $ and $x_{0}=\pm 15$ are
positions of centers of the two pulses. As seen in Eqs.~(\ref{icc1})--(\ref%
{icc2}), the soliton components are lent opposite initial momenta and, as a
result, they propagate in opposite directions, as shown in the top panel of
Fig.~\ref{fig1}. We stress that even though a small amount of radiation is
emitted (see the four bottom panels of Fig.~\ref{fig1}), the two dark
solitons in the $\psi _{\pm 1}$ fields, coupled to their bright counterparts
in the $\psi _{0}$ component, propagate practically undistorted, and around $%
t=19$ they undergo a \textit{quasi-elastic} collision; moreover, it is
clearly observed that the solitons remain unscathed after the collision.
This result is consistent with our asymptotic calculations performed above,
indicating that the small-amplitude limit (for sufficiently small $\delta $)
of the nonintegrable system of Eqs.~(\ref{dvgp1})--(\ref{dvgp2}) behaves
like the integrable YO system.

\begin{figure}[tbp]
~ \includegraphics[width=4.15cm,height=2.9cm]{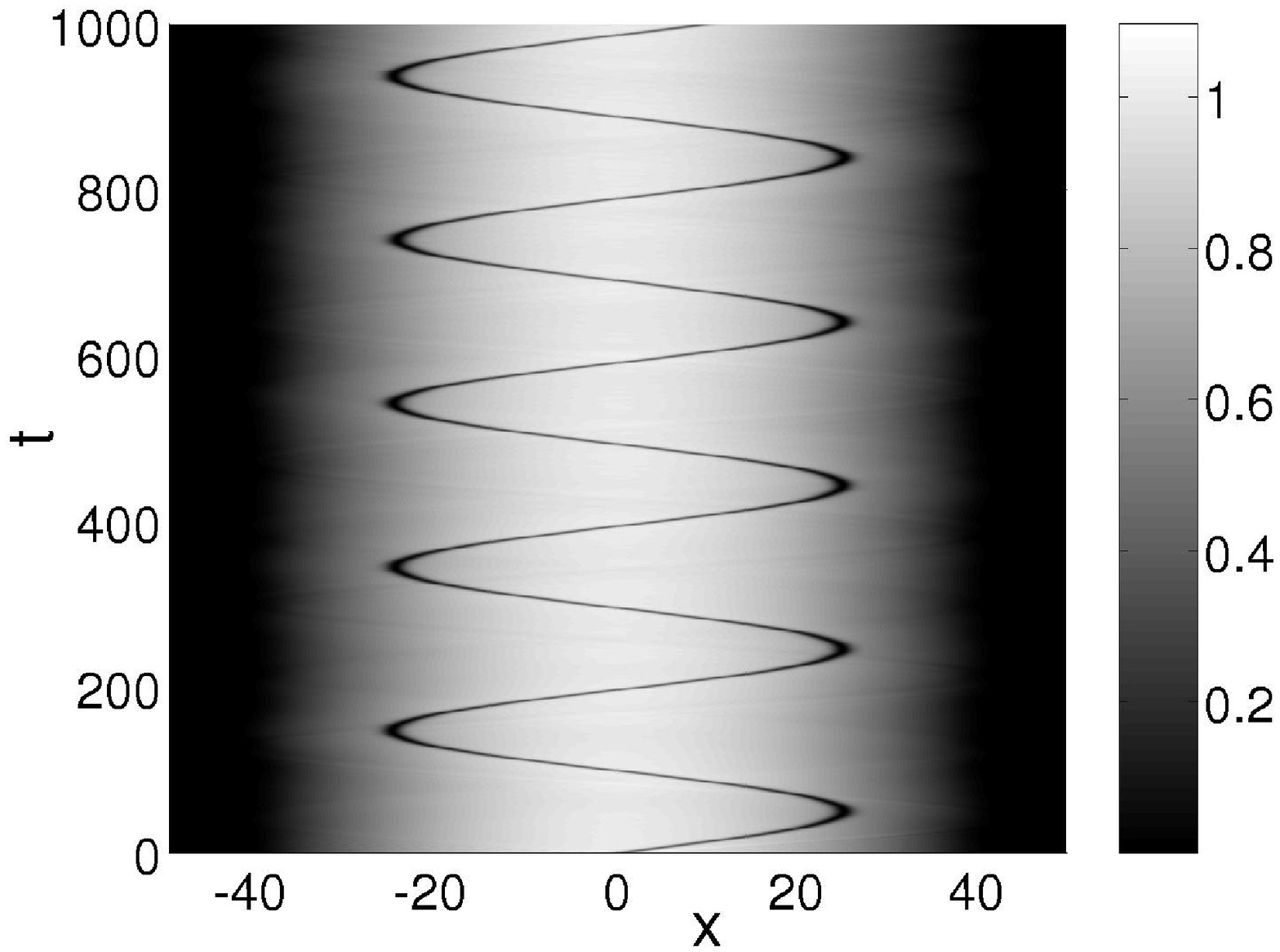} %
\includegraphics[width=4.15cm,height=2.9cm]{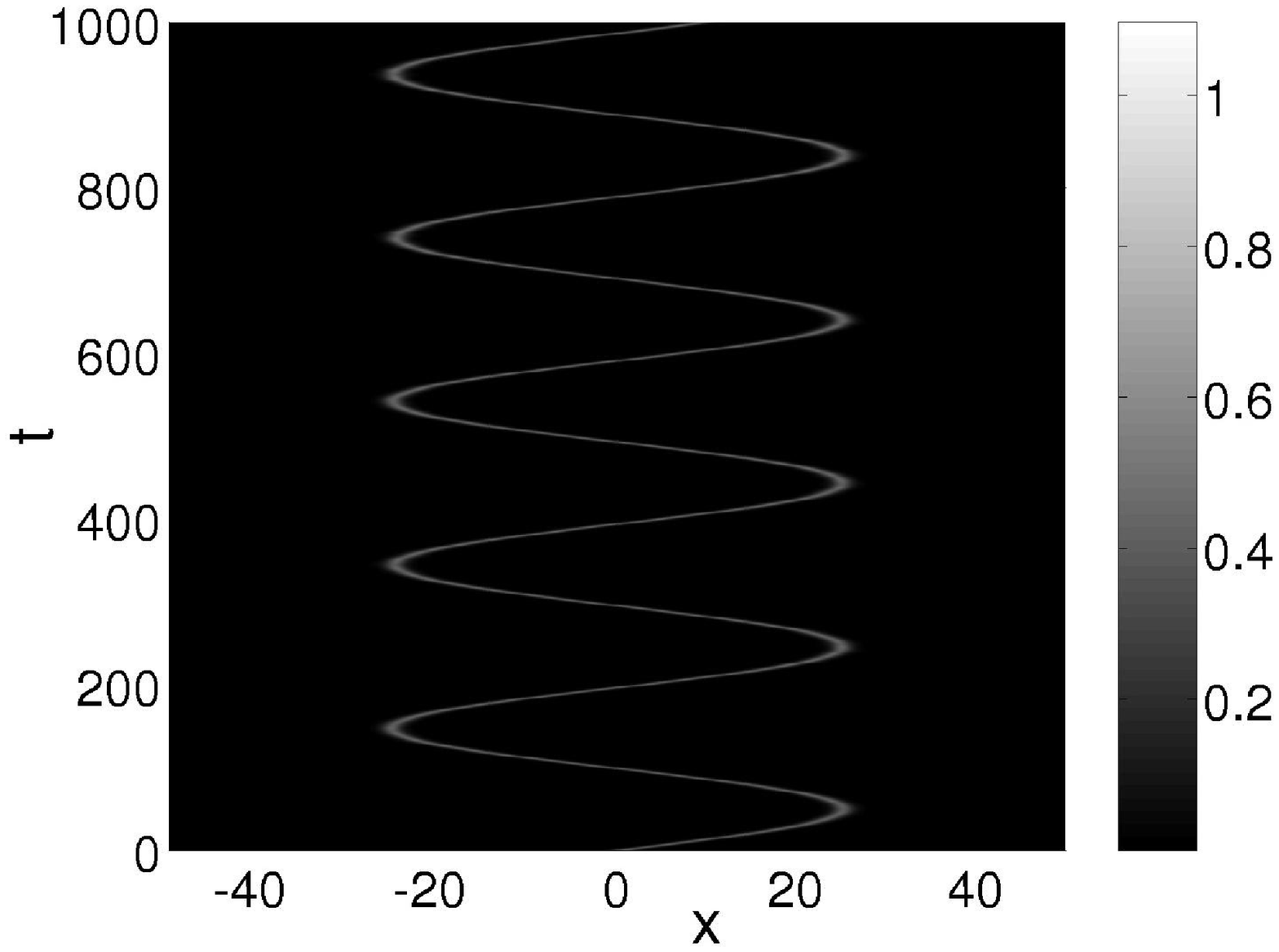}\\[1.0ex]
\includegraphics[width=3.95cm,height=2.9cm]{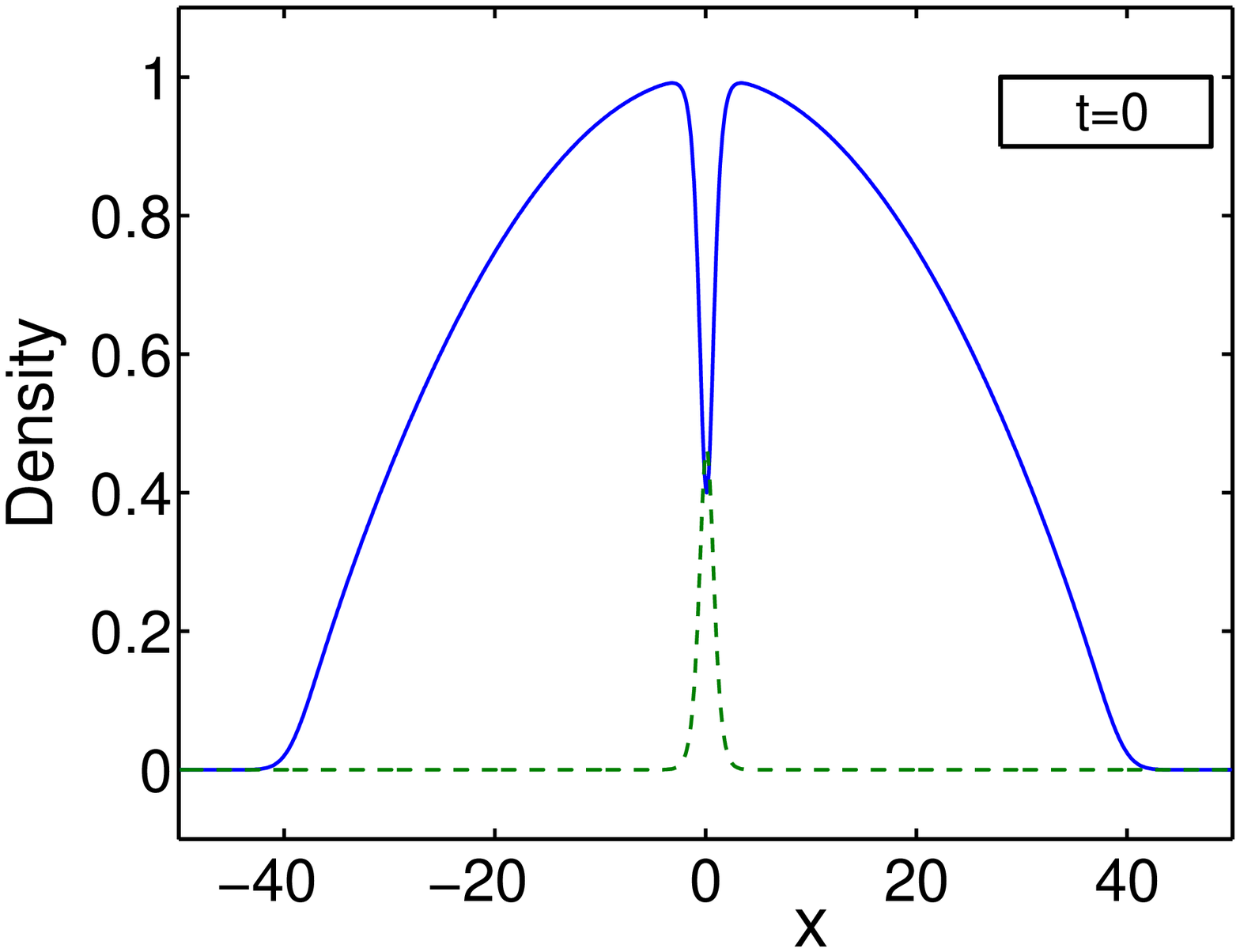} ~ %
\includegraphics[width=3.95cm,height=2.9cm]{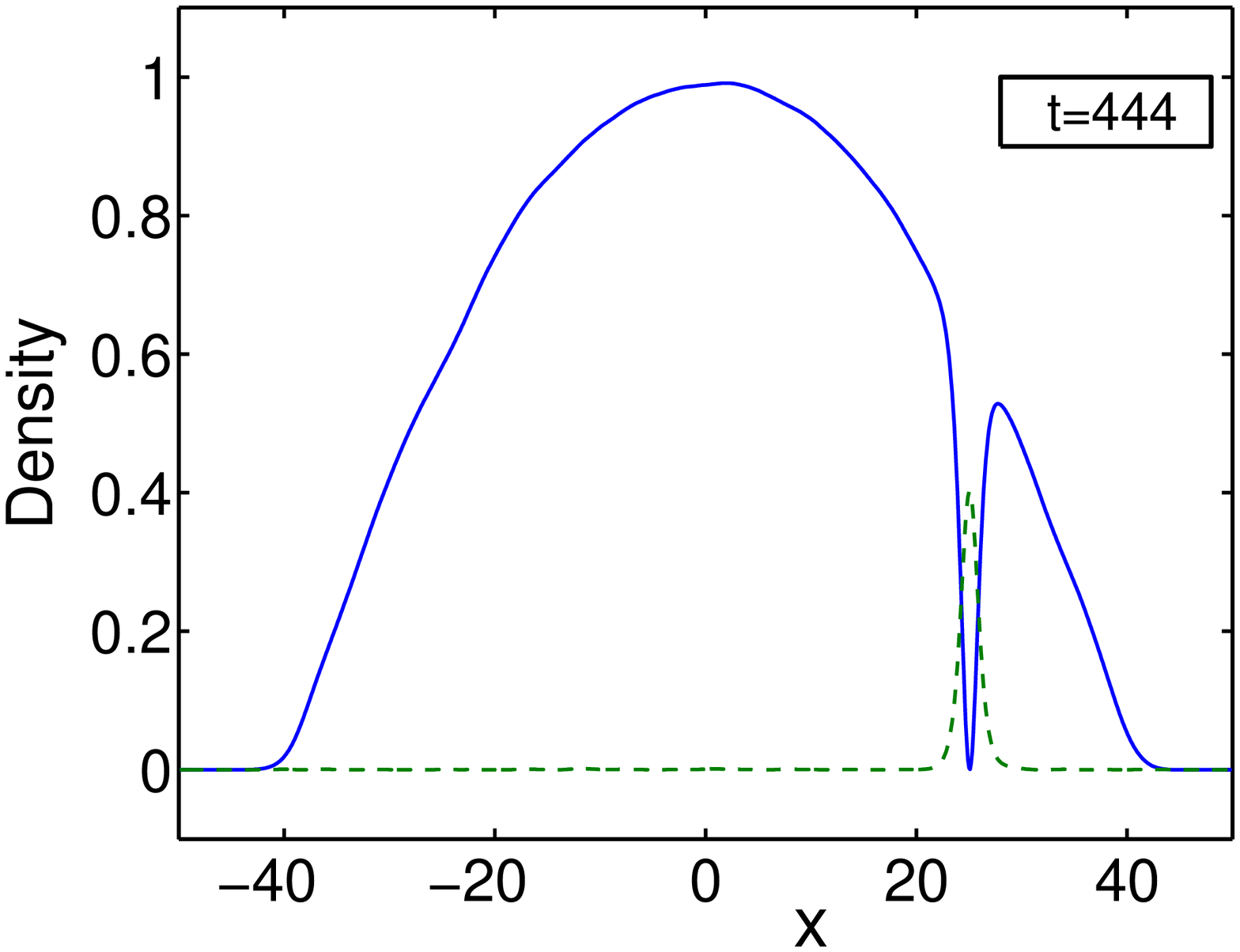}\\[1.0ex]
\includegraphics[width=3.95cm,height=2.9cm]{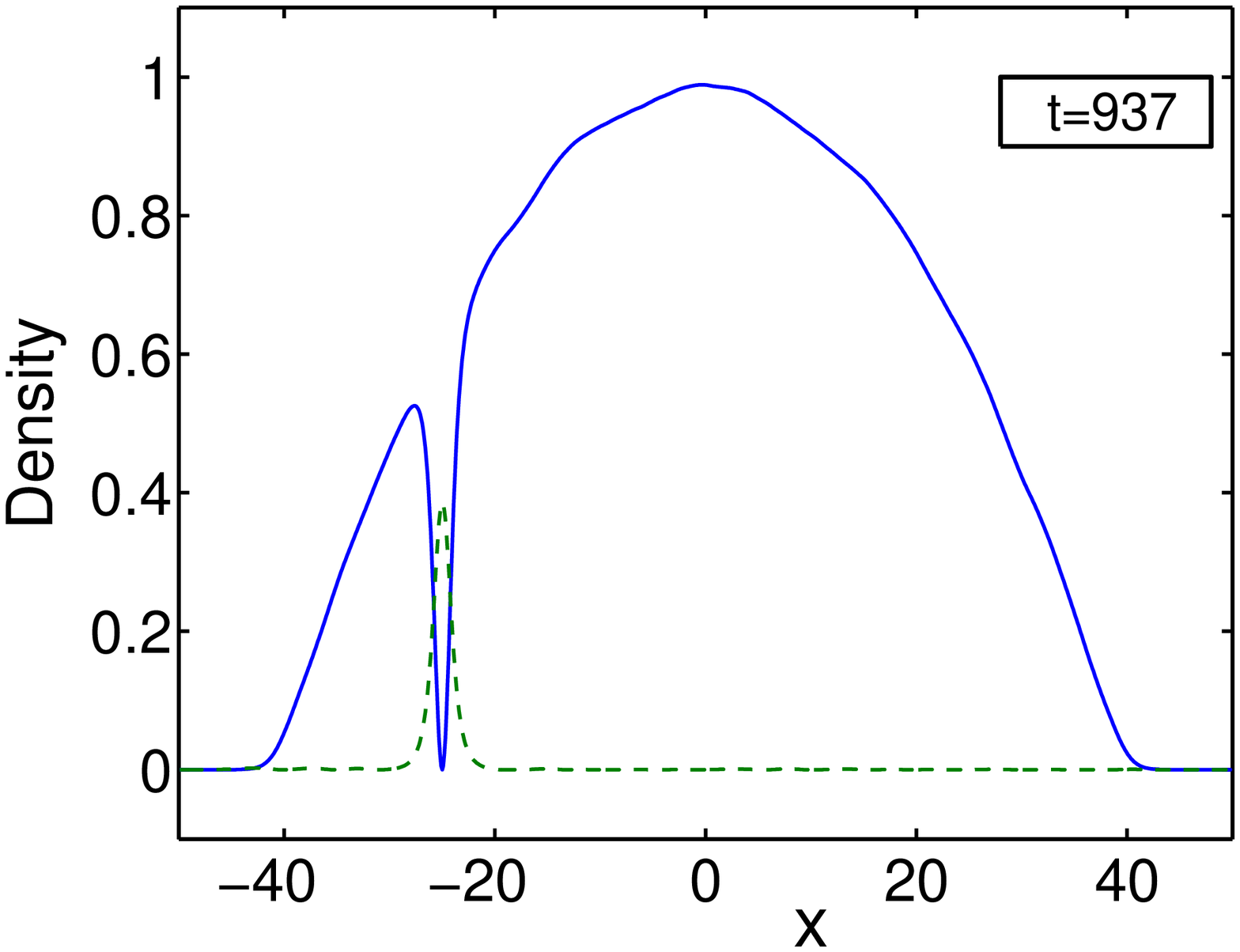} ~ %
\includegraphics[width=3.95cm]{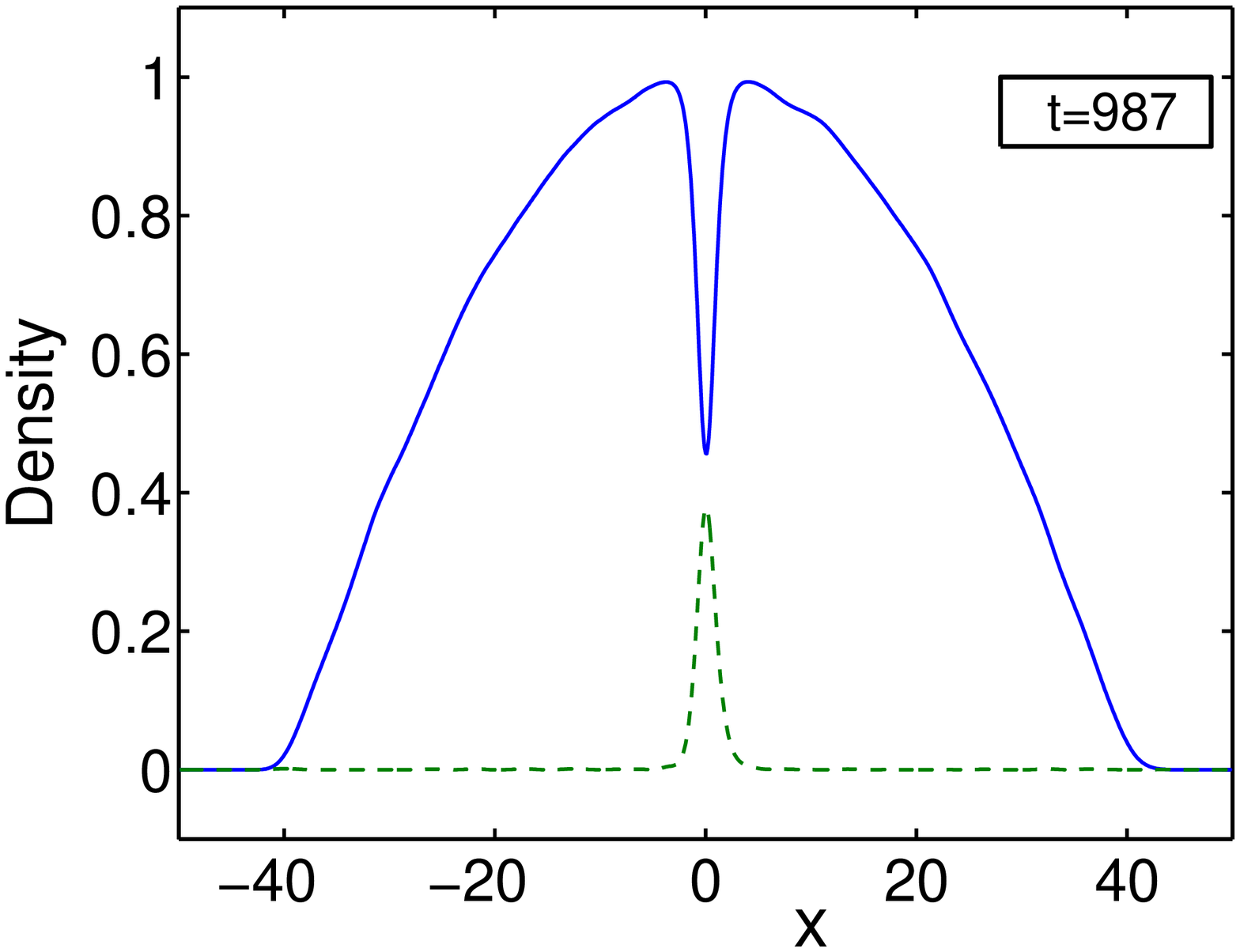}
\vspace{-0.4cm}
\caption{ The two top panels show contour plots of the densities of the $%
\protect\psi _{\pm 1}$ (left) and $\protect\psi _{0}$ (right) fields
confined in the harmonic trap with $\Omega _{\mathrm{tr}}=0.05$ (the other
parameters are the same as in Fig.~\protect\ref{fig1}). Initially, each of
the Thomas-Fermi profiles of the $\protect\psi _{\pm 1}$ components carries
a dark soliton, while the $\protect\psi _{0}$ component is a bright soliton
(the initial position is at the trap's center, $x=0$). The four bottom
panels show snapshots of the densities observed at $t=0,~444,~937,$ and $987$%
. }
\label{fig2}
\end{figure}

Next, we consider the confined system, with $\Omega _{\mathrm{tr}}=0.05$. In
this case, strictly speaking, the asymptotic reduction of the primary system
of Eqs.~(\ref{dvgp1})--(\ref{dvgp2}) to the YO system [Eqs.~(\ref{yo1}) and (%
\ref{yo2})] is not valid. However, even in the presence of the potential
term, the solutions obtained with $\Omega _{\mathrm{tr}}=0$ may be used as
an initial configuration set near the bottom of the trap, to generate DDB-
(or BBD-) like solutions of the inhomogeneous system. To that end, we first
integrate Eqs.~(\ref{dvgp1})--(\ref{dvgp2}) in imaginary time, finding a
ground state of the Thomas-Fermi (TF) type for fields $\psi _{\pm 1}$, which
is approximated by the well-known analytical density profile \cite{pit}, $%
n_{\pm 1}=(1/2)(\mu -\Omega _{\mathrm{tr}}^{2}x^{2})$. Then, at $t=0$, the
initial conditions for the $\psi _{\pm 1}$ components are taken as the
numerically found TF profiles multiplied by the dark soliton as in Eq.~(\ref%
{ic1}), while the initial configuration of the $\psi _{0}$ field is taken as
the bright soliton in Eq.~(\ref{ic2}).

In such a case, and given that the spinor DDB solitons were found above to
be robust objects behaving similar to genuine solitons of an integrable
system, one would expect that the solitons would perform harmonic
oscillations in the presence of the (sufficiently weak) parabolic trap.
Figure \ref{fig2} shows that this is the case indeed: The DDB soliton, which
was initially placed at the trap's center, oscillates as a whole without
significant deformations of its components up to large times [while the
figure extends to $t=1000$ (which is $1.2$ seconds in physical units), a
similar behavior continues at still larger times].
This is a clear indication to the fact that the predicted DDB complexes have
a good chance to be observed in the experiment. A noteworthy feature of the
numerical data is that the bright-soliton component is guided by the dark
ones, the entire soliton complex oscillating at a single frequency. The
value of the frequency is estimated below.

\subsection{The soliton's oscillation frequency}

The effect of the harmonic trap on the spinor-soliton dynamics can be
studied analytically, using the asymptotic multiscale expansion, similar to
how it was done in Refs.~\cite{huang,tonks1,tonks2}. In those works,
asymptotic reductions of various scalar GPE-based models, which included the
trapping potential, led to Korteweg-de Vries equations with \textit{variable}
coefficients [instead of constant ones that would be the case for the
homogeneous (untrapped) system]. In the present situation, we may expect,
accordingly, that the inclusion of the harmonic potential may lead to a YO
system with variable coefficients.

An important result that can be produced by such an analysis is the
oscillation frequency of the solitons in the presence of the harmonic trap.
As shown in Refs.~\cite{huang,tonks1,tonks2} by means of the adiabatic
perturbation theory for solitons \cite{RMP}, the oscillation frequency can
be obtained from the equation (which is valid to leading order in the small
parameter characterizing the inhomogeneity of the system),
\begin{equation}
\frac{d\tilde{X}}{dt}=\tilde{c}(\tilde{X}),  \label{sc}
\end{equation}%
where $\tilde{X}$ is a properly chosen slow spatial variable, and $\tilde{c}$
is the speed of sound, which is now position-dependent due to the background
density of the trapped condensate. In fact, Eq.~(\ref{sc}) is a
straightforward generalization of the result obtained for the homogeneous
system, where it has been found that the speed of sound is $c=\sqrt{\mu }$
[see Eq.~(\ref{sound})], while the soliton's velocity is $v=\sqrt{\mu }-2%
\sqrt{\delta }\xi $, implying that (for sufficiently small $\delta $) $%
dx/dt\approx \sqrt{\mu }\equiv c$.

To adopt this approach to the present case, we need to find the local speed
of sound $\tilde{c}(\tilde{X})$ when the harmonic potential term, $%
V=(1/2)\Omega _{\mathrm{tr}}^{2}x^{2}$, is included in the spin-independent
part of the Hamiltonian, $H_{\mathrm{si}}$. Taking into regard that the
potential varies slowly on the soliton's spatial scale, $\nu ^{-1/2}$ (see,
e.g., Fig.~\ref{fig2}), we define the above-mentioned slow spatial variable
as $\tilde{X}=\tilde{\epsilon}x$, where $\tilde{\epsilon}=\Omega _{\mathrm{tr%
}}/\tilde{\Omega}_{\mathrm{tr}}$ [recall that $\Omega _{\mathrm{tr}}$ given
in Eq.~(\ref{Omegatr}) is of order $10^{-2}$], and $\tilde{\Omega}_{\mathrm{%
tr}}$ is an auxiliary $O(1)$ scale parameter. This way, the trapping
potential takes the form of $V(\tilde{X})=(1/2)\tilde{\Omega}_{\mathrm{tr}%
}^{2}\tilde{X}^{2}$, i.e., it depends only on slow variable $\tilde{X}$.

Then, the local speed of sound can easily be derived upon considering the
linearization of Eqs.~(\ref{h1}) and (\ref{h2}), which are modified by the
inclusion of the term $-nV(\tilde{X})$ in the left-hand side of Eq.~(\ref{h1}%
). The ground state of this system can easily be found by setting $v\equiv
\phi _{x}=0$ and $\phi _{t}=-\mu $. Then, since Eq.~(\ref{h1}) implies that $%
n=n_{0}$ is time-independent in the ground state, we assume that $%
n_{0}=n_{0}(\tilde{X})$ and, to the leading order in $\tilde{\epsilon}$, we
obtain
\begin{equation}
n_{0}(\tilde{X})=(1/2)\left[ \mu -V(\tilde{X})\right] ,  \label{TF}
\end{equation}%
in the region where $\mu >V(\tilde{X})$, and $n_{0}=0$ outside. Equation (%
\ref{TF}), which is the TF approximation for the density profile, also
implies that, for $V(\tilde{X})=(1/2)\tilde{\Omega}_{\mathrm{tr}}^{2}\tilde{X%
}^{2}$ the axial size of the trapped condensate is $2L\equiv 2\sqrt{2\mu }%
/\Omega $. Similarly to the analysis presented above in Sec.~\ref{SEC:modelB}, 
we now
consider the linearization around the ground state and seek for respective
solutions to Eqs.~(\ref{h1})--(\ref{h2}) as $n=n_{0}(\tilde{X})+\tilde{%
\epsilon}n_{1}(x,t)$, $\phi =-\mu _{0}t+\tilde{\epsilon}\phi _{1}(x,t)$, and
$\Phi _{0}=\tilde{\epsilon}\Phi _{1}(x,t)$, with $n_{1},\phi _{1},\Phi
_{1}\sim \exp [i(kx-\omega t)]$. This way, we obtain the following
dispersion relation for the inhomogeneous system,
\begin{equation}
\omega ^{2}=2n_{0}(\tilde{X})k^{2}+(1/4)k^{4},  \label{mdr}
\end{equation}%
%
%
%
and, accordingly, the local speed of sound:
\begin{equation}
c(\tilde{X})=\sqrt{2n_{0}(\tilde{X})},  \label{cs}
\end{equation}%
which bears resemblance to the sound propagation in weakly nonuniform media
\cite{landau}; in the homogeneous case, Eq.~(\ref{cs}) is reduced to Eq.~(%
\ref{sound}).

Next, we substitute Eq.~(\ref{cs}) in Eq.~(\ref{sc}) and, taking into regard
the density profile given by Eq.~(\ref{TF}), we integrate the resulting
first-order differential equation. The result is
\begin{equation}
\tilde{X}=L\sin \left[ \left( \tilde{\Omega}_{\mathrm{tr}}/\sqrt{2}\right)
t\right] ,  \label{m}
\end{equation}%
which demonstrates that a sufficiently shallow spinor dark soliton
oscillates with frequency 
\begin{equation}
\omega_{\rm osc} = \frac{\Omega_{\rm tr}}{\sqrt{2}}, 
\label{op}
\end{equation}
similarly to the known result for the oscillations of dark solitons in a
single-component BEC \cite{oscfreq} [note that, for the sake of simplicity, 
we dropped the tilde in Eq.~(\ref{op})]. 
\begin{figure}[tbp]
 \includegraphics[width=4.37cm,height=3.50cm]{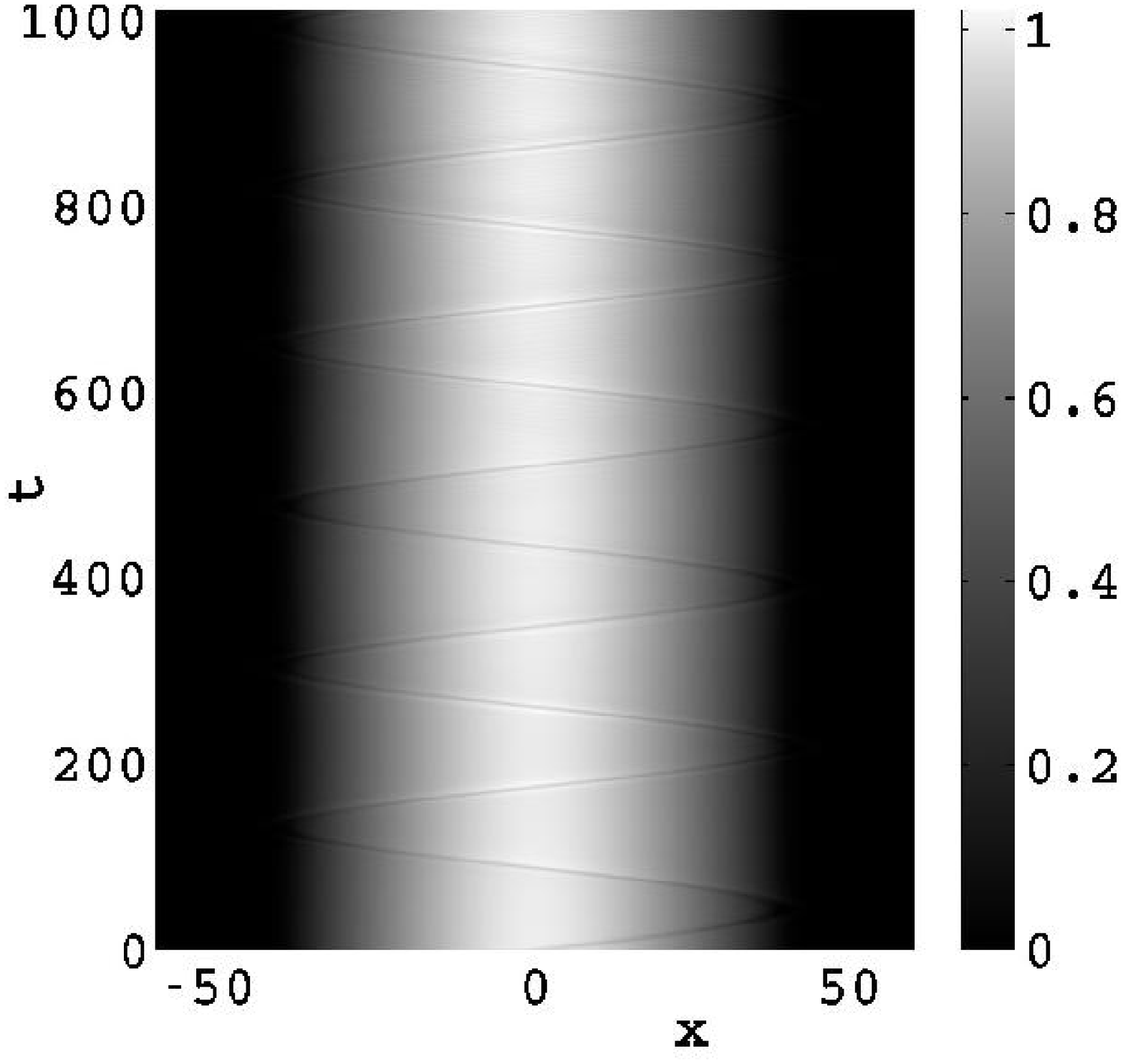} %
~~\includegraphics[width=3.93cm,height=3.60cm]{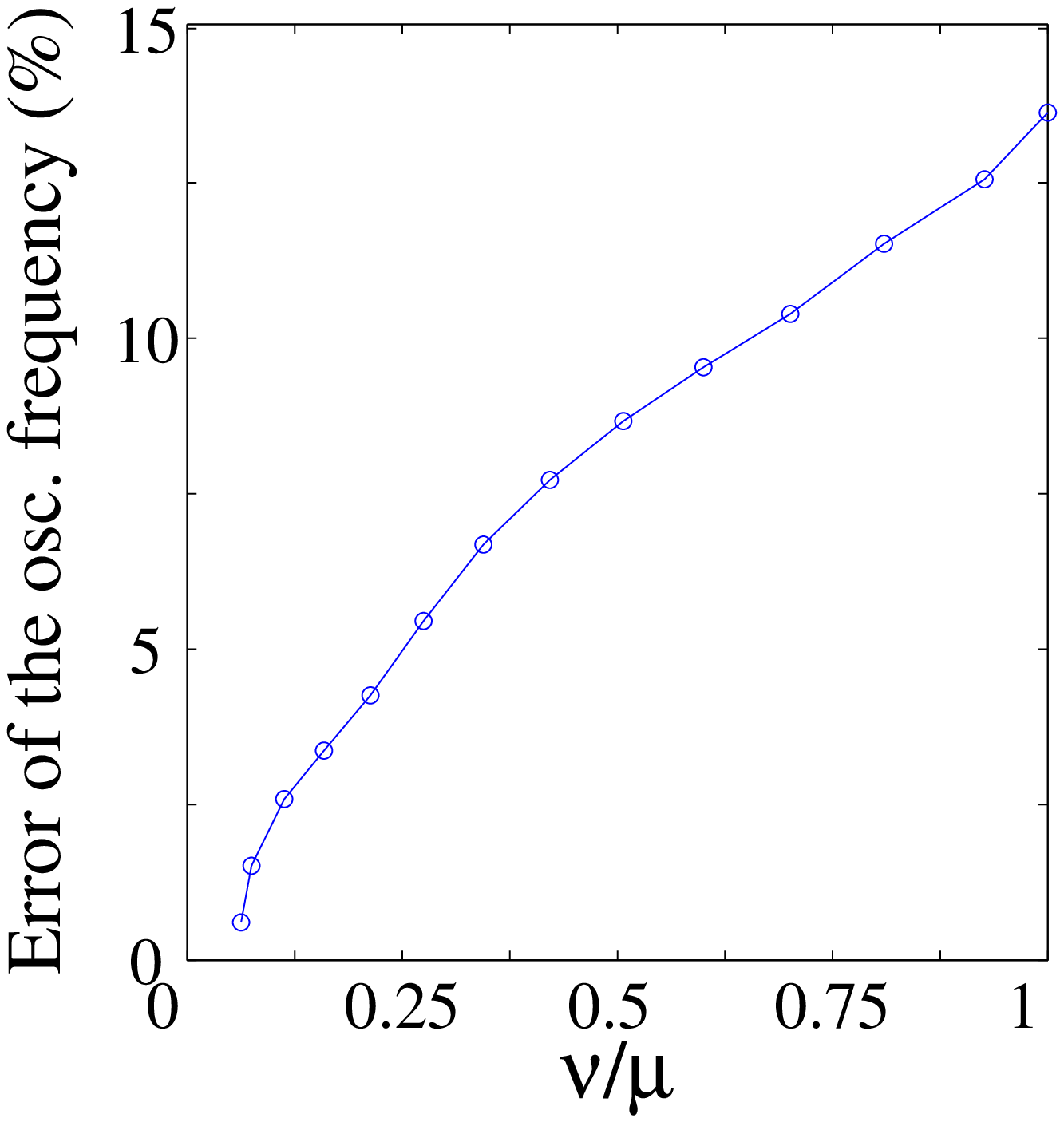}\\[-1.0ex]
\vspace{-0.4cm}
\caption{
Left panel: Contour plot of the density of a shallow dark soliton (shown are 
the identical $|\protect\psi _{\pm 1}|^{2}$ fields) with initial amplitude $\nu=0.13$; 
the other parameter values are $\eta=1$, $\xi=0.5$, while the values of the chemical potential 
and the harmonic trap strength are $\mu=2$ and $\Omega _{\mathrm{tr}}=0.05$ as before. 
The soliton performs oscillations with a frequency $\omega_{\mathrm{osc}}\approx 0.03513$, 
which is almost identical to the analytical prediction of $\Omega_{\mathrm{tr}}/\sqrt{2}= 0.03536$ (error $\approx 0.65\%$).
Right panel: Percentage error of the numerically found soliton oscillation frequency, as compared to the analytical prediction 
of Eq.~(\ref{op}), as a function of the relative dark soliton depth $\nu/\mu$. The region 
$\nu/\mu \ll 1$ corresponds to shallow solitons, while $\nu/\mu = 1$ corresponds to a black soliton; 
for the case shown in the left panel $\nu/\mu=0.065$. As seen, 
in the former case the analytical prediction is fairly good (the error is below $5\%$ for every $\nu/\mu <0.25$), while 
for large amplitude solitons $\nu/\mu > 0.7$ the error becomes more than $10\%$, with a maximum $13.7\%$ for $\nu/\mu=1$.
}
\label{fig3}
\end{figure}
We should clearly stress here that we expect the above result 
to be valid in the case of shallow solitons, with a relative depth $\nu/\mu \ll 1$ [see Eq.~(\ref{ic1})]. 
For example, as seen in the left panel of Fig.~\ref{fig3}, in the case of such a shallow soliton with $\nu/\mu =0.065$ (corresponding to the 
above mentioned physically relevant choice of $\nu = 0.13$ and a value of the chemical potential $\mu=2$) the analytical 
prediction is quite accurate: the numerically found oscillation frequency (for $\Omega_{\mathrm{tr}}=0.05$) is $\approx 0.03513$, 
while the analytical prediction of Eq.~(\ref{op}) is $0.03536$ (the percentage error is $\approx 0.65\%$). 
On the other hand, as seen in the same figure, the amplitude of the soliton oscillation is $39.992$, while 
$L=\sqrt{2\mu }/\Omega = 40$ (the respective percentage error is $0.02\%$. 

Although the result of Eq.~(\ref{m}) [and (\ref{op})] is ``compatible'' with our analytical approach developed in the previous section, which is 
also valid for small-amplitude solitons, it is  also of interest 
to numerically investigate the oscillations of solitons of moderate 
and large amplitudes. In this respect, in the right panel of Fig.~\ref{fig3} we show 
the percentage error of the numerically found soliton oscillation frequency [relative to the prediction of Eq.~(\ref{op})] 
as a function of the relative dark soliton depth $\nu/\mu$. It is clear that the region 
$\nu/\mu \ll 1$ corresponds to shallow solitons, while the limiting case of $\nu/\mu = 1$ corresponds to a ``black'' soliton, 
with an initially zero intensity at the trap's center (the latter is slightly displaced from the trap's center to be set into motion). 
As seen in the figure, the analytical prediction of Eq.~(\ref{op}) is very good as concerns the estimation of the oscillation 
frequency, since the error is below $2\%$ for every $\nu/\mu <0.1$. 
Notice that we have also computed the error in the oscillation amplitude
(not shown here), which we have found to be larger 
(i.e., up to $8\%$ in this regime); this is due to the fact that the 
increasingly deeper solitons are not reflected at the rims of the condensate, but
rather inside the cloud, as can be seen e.g., in Fig.~\ref{fig2}. 

In the case of moderate (and large) amplitude solitons the analytical prediction becomes worse: For example, 
in the particular case shown in Fig.~\ref{fig2} (in this case $\nu/\mu=0.6$), comparing the
numerically found soliton's oscillation frequency, $\omega_{\mathrm{osc}}\approx 0.032$ to the aforementioned prediction of 
$0.03536$ (again for $\Omega_{\mathrm{tr}}=0.05$), we find a relatively large
discrepancy ($\approx 9.5\%$) between the two values. However, an important observation regarding 
Fig.~\ref{fig2}, is that the bright-soliton component performs oscillations at the same frequency, 
$\Omega _{\mathrm{tr}}/\sqrt{2}$. This is a clear indication that the bright component is
guided (effectively trapped) by the dark component of the DDB complex. Note that, in the single-component 
BEC, bright solitons oscillate in the parabolic potential with frequency, $\Omega _{\mathrm{tr}}$ \cite{st} (as a
consequence of the Kohn's theorem \cite{kohn}), rather than $\Omega_{\mathrm{tr}}/\sqrt{2}$.

Furthermore, we observe that, naturally, the discrepancy becomes even larger 
in the case of large-amplitude (nearly-black) solitons, which perform small-amplitude oscillations around the trap's center. 
For example, for $\nu/\mu=0.8$ ($\nu/\mu=0.9$) the numerically found values of the oscillation frequency deviate from the value 
of $\Omega_{\mathrm{tr}}/\sqrt{2}$ by $11.5\%$ ($12.4\%$), while in the limiting case of $\nu/\mu=1$ the respective percentage 
error is $13.7\%$. It is clear that such large deviations are due to the fact that the above numerical results pertain to 
non-small amplitude solitons with large values of $\nu/\mu$, 
contrary to what is the case for the the analytical approach
which requires $\nu/\mu \ll 1$, as mentioned above. 

In such a case of large amplitude solitons, it is interesting to compare the numerically found oscillation frequencies to a different 
analytical prediction presented in Ref.~\cite{BA}. In this work, the oscillation frequency may result from a nonlinear equation of motion 
for the bright-dark soliton in a binary BEC mixture (see Eq.~(5) of \cite{BA}). In fact, for shallow solitons the oscillation frequency 
is identical to the one given by Eq.~(\ref{op}), while, in the opposite limit of very deep dark solitons, may be 
approximated (in our language) as,
\begin{equation}
\omega_{\rm osc} = \frac{\Omega_{\rm tr}}{\sqrt{2}} \left( 1- \frac{N_{\rm B}}{4\sqrt{\mu+(N_{\rm B}/4)^{2}}} \right)^{1/2}, 
\label{BAp}
\end{equation}
where $N_{\rm B}$ is the number of atoms of the bright-soliton component, which, employing Eq.~(\ref{ic2}), is easily found to be 
$N_{\rm B}= 2\nu^{3/2}\xi/\eta\sqrt{\mu}$. It is clear that Eq.~(\ref{BAp}) shows that the oscillation frequency is 
down-shifted (as compared to the value of $\Omega_{\rm tr}/\sqrt{2}$)
[i.e., the dark-bright pair executes slower oscillations, as the bright
component is enhanced], thus having a quantitative agreement with our 
numerical observations. 
Perhaps more importantly, it also provides better quantitative estimates for the values of the soliton's oscillation frequencies. 
In particular, for normalized soliton depths $\nu/\mu=0.8$, $0.9$ and $1$, the predictions of Eq.~(\ref{BAp}) deviate from the respective 
numerically found oscillation frequencies with percentage errors $1.7\%$, $4\%$ and $5.2\%$ 
(recall that a respective comparison with the prediction of Eq.~(\ref{op}) led to percentage errors $11.5\%$, $12.4\%$, and $13.7\%$, respectively). 
The above results indicate that a more detailed description of the motion of the bright-dark soliton complexes in the 
trapped spinor condensate, in the lines of Ref.~\cite{BA}, would be very interesting. However, this is beyond the 
scope of the present work.

\subsection{Effects of larger spin-dependent interactions}

In the previous subsections we had kept the value of the parameter $\delta$ fixed, i.e., $\delta =0.0314$ 
(corresponding to the polar spin-1 sodium BEC). Such a small value of $\delta$ validates our perturbative approach, 
which allowed us to find approximate DDB soliton solutions of the YO-type, and study their oscillations in the trapped 
spinor condensate. It is interesting, however, to test the stability and the dynamics of the obtained DDB solitons 
in the more general case of a stronger perturbation, considering also non-small values of $\delta$. 
In this respect, we will now present numerical results obtained by direct numerical integration of 
Eqs.~(\ref{dvgp1}) and (\ref{dvgp2}) for $\delta = 0.2$, which is an order of magnitude greater than the previous value. 
Note that we will study the evolution of DDB solitons with the same amplitudes as in the case of small $\delta$, 
so as to directly compare the results pertaining to weak or moderate spin-dependent interaction strength.
 
\begin{figure}[tbp]
~ \includegraphics[width=4.15cm,height=3.2cm]{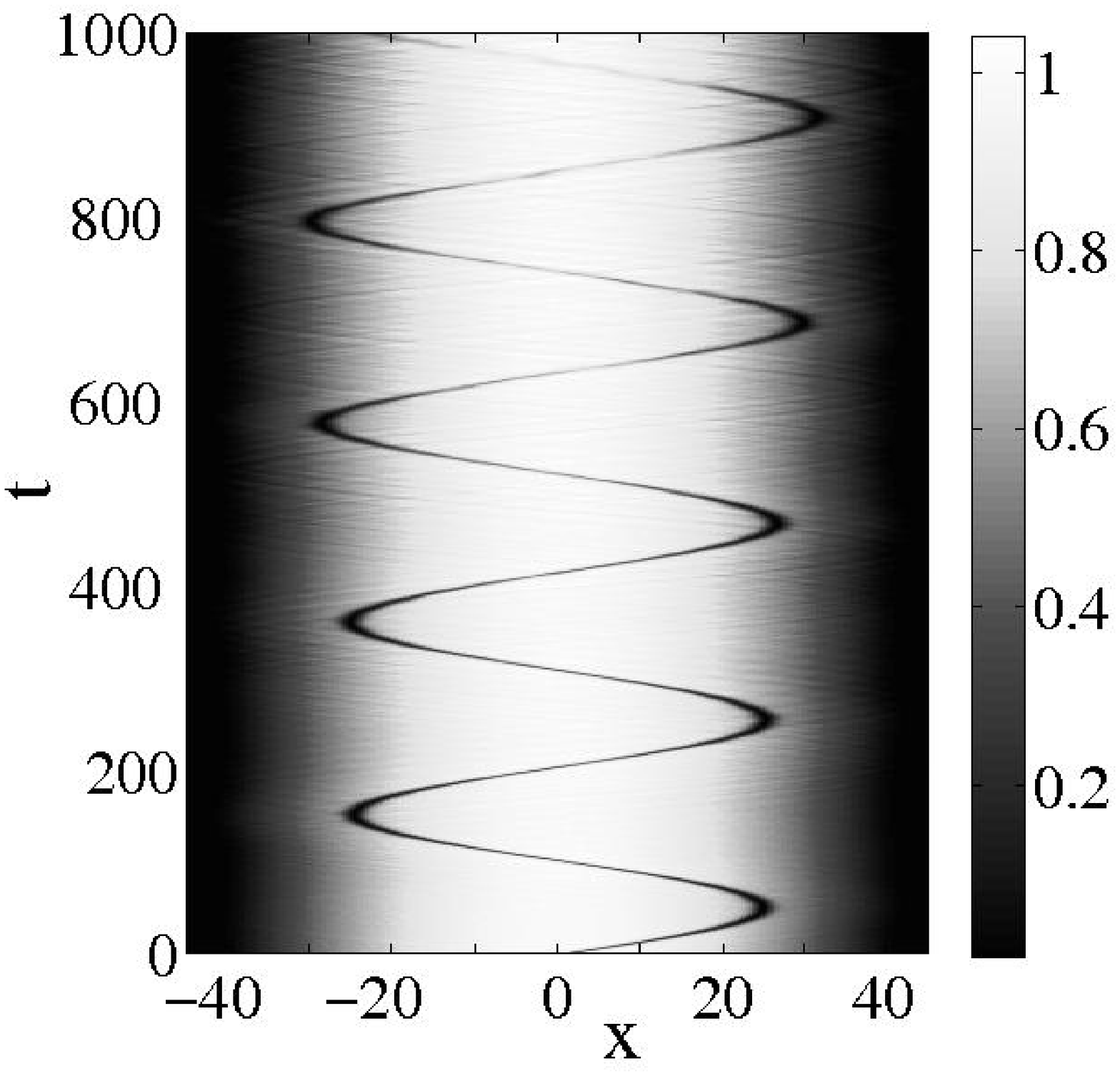} %
\includegraphics[width=4.15cm,height=3.2cm]{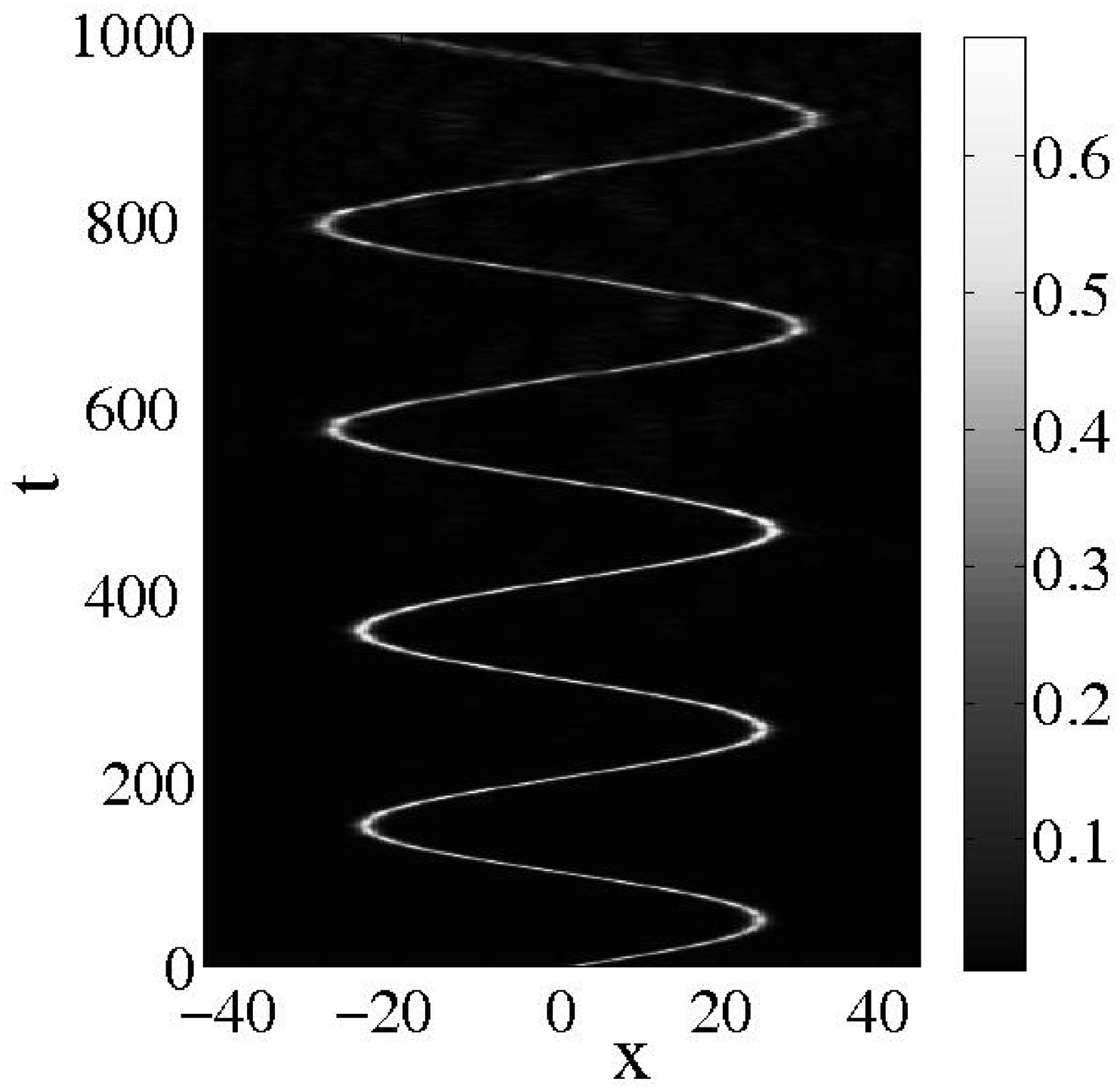}\\[2.0ex]
\includegraphics[width=3.95cm,height=3.3cm]{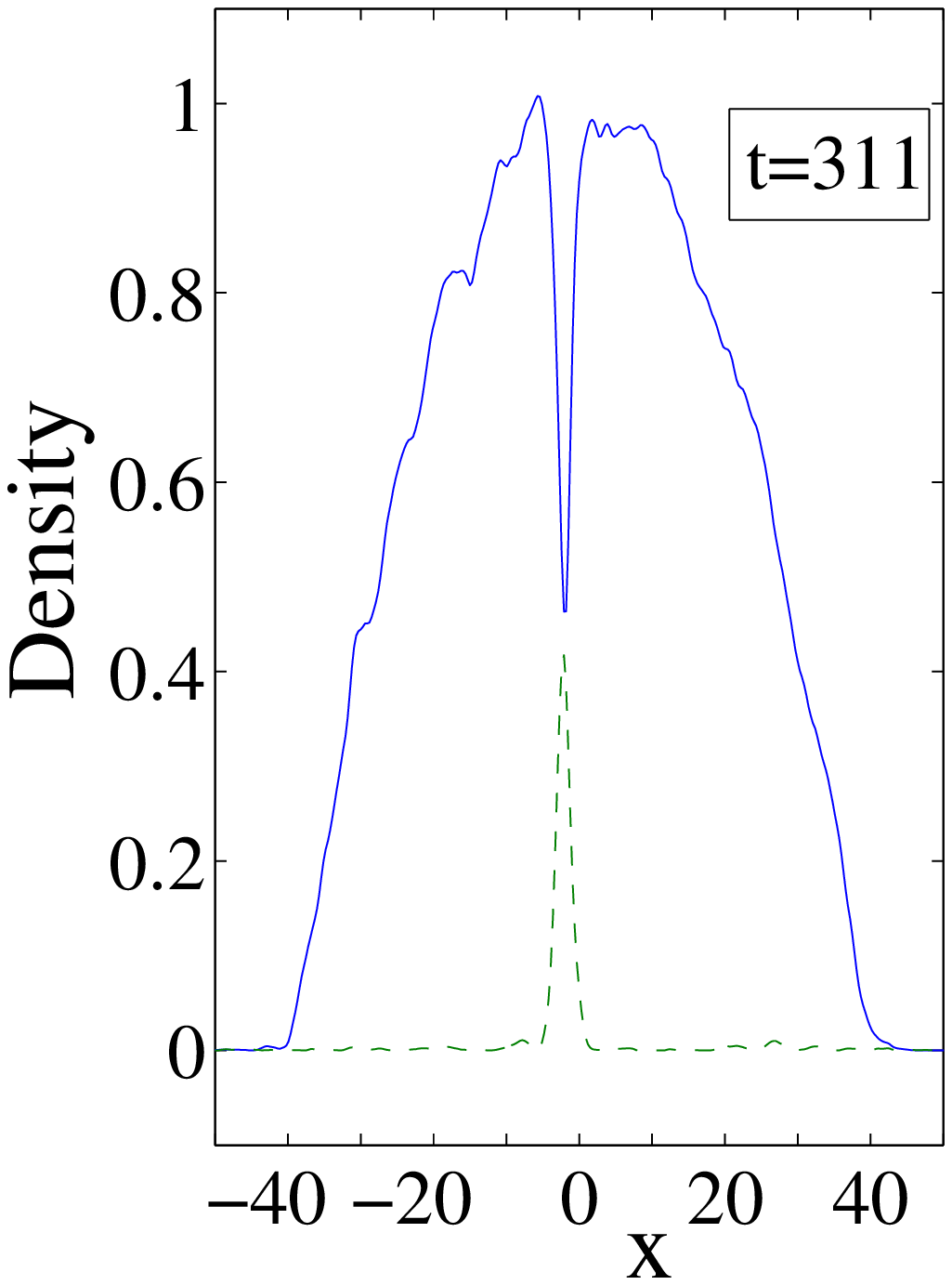} ~ %
\includegraphics[width=3.95cm,height=3.3cm]{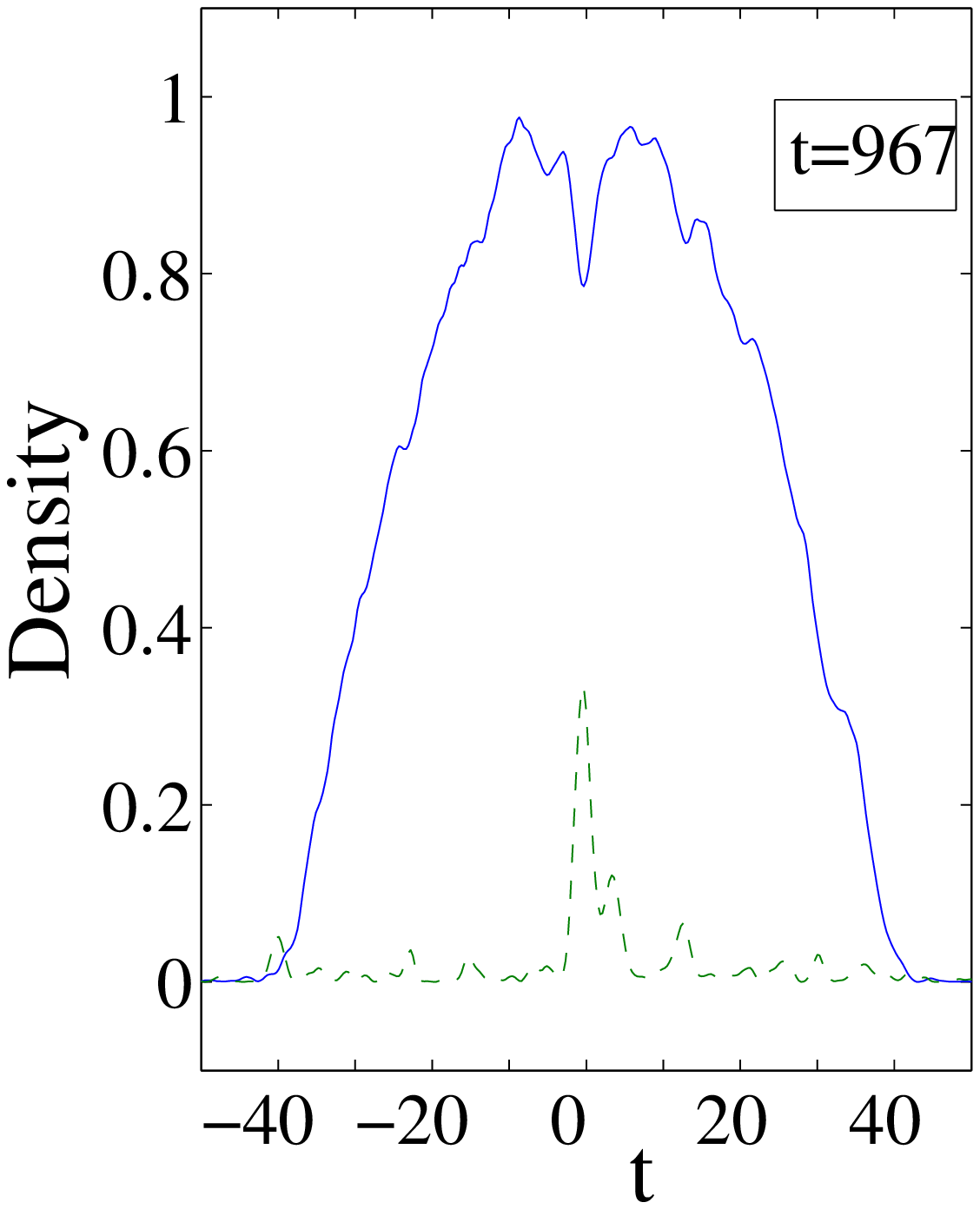}\\[2.0ex]
\vspace{-0.4cm}
\vspace{-0.2cm}
\caption{ 
Same as Fig.~\ref{fig2}, but for $\delta=0.2$. 
The soliton parameter values are $\protect \xi =0.61$, $\protect \eta =1.22$, and $\protect\nu =1.2$, while 
the values of the trap strength and chemical potential are $\Omega _{\mathrm{tr}}=0.05$ and $\protect\mu =2$. 
This choice renders the initial soliton densities identical to the ones shown in the second-row left panel of Fig.~\ref{fig2}. 
The two bottom panels show snapshots of the densities at $t=311$ and $t=967$. 
}
\label{fig4}
\end{figure}

In Fig.~\ref{fig4}, we show the evolution of a moderate amplitude DDB soliton, characterized by 
the parameters $\protect \xi =0.61$, $\protect \eta =1.22$, and $\protect\nu =1.2$; for the same values of the 
trap strength and chemical potential as before ($\Omega _{\mathrm{tr}}=0.05$ and $\protect\mu =2$), the 
initial condition is the same as the one shown in the second-row left panel of Fig.~\protect\ref{fig2} 
(in this case, recall that the normalized amplitude of the dark solitons in the $\protect\psi_{\pm 1}$ components is $\nu/\mu=0.6$).  
As observed in Fig.~\ref{fig4}, although the stronger perturbation induces emission of stronger radiation
(see bottom left panel), the losses are not significant up to relatively large times, of order of 
$t=311$ (or $t=360$ ms in physical units): the dark (bright) soliton densities 
are only $8\%$ ($9\%$) smaller than their initial values, and it can safely be concluded that even for such a strong value of 
$\delta$ the DDB vector soliton has a good chance to be observed in an experiment (provided, of course, that such a magnitude of the spin-dependent 
interaction is experimentally achieved). However, for later times the 
continuous perturbation-induced emission of radiation results in the eventual destruction of the DDB complex. 
In particular, at $t \approx 1000$, the solitons almost decay: at $t=967$ (see bottom right panel of Fig.~\ref{fig4}), 
the density of the dark (bright) soliton is $63\%$ ($28\%$) smaller than its initial value.

The cases of large- and small-amplitude solitons, with normalized amplitudes (of the dark soliton) 
being $\nu/\mu=0.8$ and $\nu/\mu=0.065$, respectively, was examined too (results not shown here). 
It was found that, naturally, the large-amplitude DDB soliton starts to accelerate immediately due to the strong radiation 
emitted and decays fast, being completely destroyed at $t \approx 300$:  
the soliton densities are smaller than $50\%$ of their initial values. 
On the other hand, the small-amplitude DDB soliton was 
found to be slightly more robust, showing a behavior similar to the one of the moderate amplitude soliton shown 
in Fig.~\ref{fig4}, but for smaller times: in fact, the vector soliton persists up to $t \approx 600$ 
(as per the above criterion, the densities are smaller than $50\%$ of their initial values) and then decays. 

To conclude this subsection, it is clear that the small- and moderate-amplitude YO-type DDB vector soliton 
persist in the spin-1 condensate up to experimentally relevant times even for large spin-dependent interaction strength, namely  
an order of magnitude larger than the physically relevant value pertaining to the polar sodium spinor condensate.

\section{Conclusions\label{SEC:conclu}}

We have studied bright-dark soliton complexes in polar spinor Bose-Einstein
condensates, both analytically and numerically. Our analytical approach is
based on the small amplitude-asymptotic reduction of the nonintegrable
vector (three-component) system of the coupled Gross-Pitaevskii equations to
a completely integrable model, \textit{viz}., the Yajima-Oikawa (YO) system.
Borrowing soliton solutions of the YO system and inverting the reduction, we
have obtained an analytical approximation for small-amplitude vector
solitons of the dark-dark-bright and bright-bright-dark types, in terms of
the $m_{F}=+1,-1,0$ components, respectively. The analytical predictions
were confirmed by direct numerical simulations. The so constructed
approximate soliton states were found to propagate undistorted and undergo
quasi-elastic collisions, featuring properties of genuine solitons.

The effect of the harmonic trapping potential (which also contributes toward
the nonintegrability of the underlying equations) on the solitons was also
studied numerically and analytically. It was found that even vector solitons
of moderate (non-small) amplitudes maintain their identity in the presence
of the parabolic trap, and perform harmonic oscillations, at large times ($%
\sim 10$ seconds or even more, in physical units). We have found that the
soliton's oscillation frequency takes (in the analytical approximation) the
value of $\Omega _{\mathrm{tr}}/\sqrt{2}$, the same as featured by the dark
soliton in the single-component repulsive trapped condensate. The numerical
results verify the analytical prediction, for sufficiently shallow dark solitons: 
It was found that for initial soliton depths below $10\%$ of the chemical 
potential, the deviation of the analytical prediction from 
the numerically found oscillation frequency was below $2\%$ 
(the error in the estimation of the amplitude of the soliton 
oscillation was below $8\%$). For moderate and large amplitude 
solitons, the discrepancy in the frequency 
was larger (of the order of more than $10\%$); 
however, in the case of very deep dark solitons we found that the respective prediction 
of Ref.~\cite{BA} led to a significantly smaller error, less than $5\%$. 
Thus, an elaborated description of the bright-dark soliton motion (for solitons of 
arbitrary amplitude) in the trapped spinor condensate is certainly a challenge for future work.

We also tested the robustness of the derived vector soliton solutions in the case of a 
large normalized spin-dependent interaction strength, namely an order of magnitude larger than the 
physically relevant one corresponding to the polar sodium spinor condensate.
We found that although the solitons are generically destroyed under such a strong perturbation, 
the lifetimes of small- and moderate-amplitude DDB solitons was more than 300 ms in physical units. 
Thus, the vector solitons obtained in this work have a good chance to be observed in an experiment.

The bright-soliton component(s) was found to be \textit{guided} by their dark counterpart(s),
oscillating with the frequency imposed by dark components. This is an
example of the all-matter-wave soliton guidance, with potential applications
in the design of quantum switches and splitters.

The work of H.E.N. and D.J.F. was partially supported by the Special
Research Account of the University of Athens. H.E.N. acknowledges partial
support from EC grants PYTHAGORAS-I. The work of B.A.M. was supported, in a
part by the Israel Science Foundation through the Center-of-Excellence grant
No. 8006/03, and German-Israel Foundation (GIF) No. 149/2006.

\end{document}